\documentclass[useAMS,usenatbib]{mnras}

\usepackage{graphicx,natbib,url,twoopt}

\usepackage{amsmath}
\usepackage{amssymb}

\usepackage{arydshln}

\bibpunct{(}{)}{;}{a}{}{,}    
\makeatletter
 \newcommandtwoopt{\citeads}[3][][]{\href{http://adsabs.harvard.edu/abs/#3}%
   {\def\hyper@linkstart##1##2{}%
    \let\hyper@linkend\@empty\citealp[#1][#2]{#3}}}    
 \newcommandtwoopt{\citepads}[3][][]{\href{http://adsabs.harvard.edu/abs/#3}%
   {\def\hyper@linkstart##1##2{}%
    \let\hyper@linkend\@empty\citep[#1][#2]{#3}}}      
 \newcommandtwoopt{\citetads}[3][][]{\href{http://adsabs.harvard.edu/abs/#3}%
   {\def\hyper@linkstart##1##2{}%
    \let\hyper@linkend\@empty\citet[#1][#2]{#3}}}      
 \newcommandtwoopt{\citeyearads}[3][][]%
   {\href{http://adsabs.harvard.edu/abs/#3}%
   {\def\hyper@linkstart##1##2{}%
    \let\hyper@linkend\@empty\citeyear[#1][#2]{#3}}}   
\makeatother

\def\lcdm{$\Lambda$CDM\ }
\def\units{cm$^{2}$g$^{-1}$}

\title[Dark matter haloes in the 2cDM model]
  {Dark matter haloes in the multicomponent model. III. From dwarfs to galaxy clusters}
\author[]
  {Keita Todoroki,$^1$\thanks{Email: keita@ku.edu}
  Mikhail V. Medvedev$^{1}$
  \newauthor 
             \\
  $^1$Department of Physics and Astronomy, University of Kansas, Lawrence, KS 66045
  }
 \date{\today}

\pagerange{\pageref{firstpage}--\pageref{lastpage}} \pubyear{2016}

\def\LaTeX{L\kern-.36em\raise.3ex\hbox{a}\kern-.15em
    T\kern-.1667em\lower.7ex\hbox{E}\kern-.125emX}

\begin{document}


\label{firstpage}

\maketitle

\begin{abstract}
A possibility of DM being multicomponent has a strong implication on resolving decades-long known cosmological problems on small scale. In addition to elastic scattering, the model allows for inelastic interactions, which can be characterized by a `velocity kick' parameter. 
The simplest 2cDM model with cross section $0.01\lesssim\sigma/m<1\textrm{ cm}^{2}{ \rm g}^{-1}$ and the kick velocity $V_{k}\simeq 100\textrm{ km s}^{-1}$ has been shown to robustly resolve the missing satellites, core-cusp, and too-big-to-fail problems in $N$-body cosmological simulations tested on MW-like haloes of a virial mass $\sim5 \times 10^{11}$ M$_{\odot}$ (Paper I $\&$ II). 
With the aim of further constraining the parameter space available for the 2cDM model, we extend our analysis to dwarf and galaxy cluster haloes with their virial mass of $\sim 10^7 - 10^8$ and $\sim 10^{13} - 10^{14}$ M$_{\odot}$, respectively. We find $\sigma_{0} / m \gtrsim 0.1$ \units is preferentially disfavored for both dwarfs and galaxy cluster haloes in comparison with observations, while $\sigma_{0} / m = 0.001$ \units causes little perceptible difference from that of the CDM counterpart for most of the cross section's velocity dependence studied in this work. Our main result is that within the reasonable set of parameters the 2cDM model can successfully explain the observational trends seen in dwarf galaxy and galaxy cluster haloes and the model leaves us an open window for other possible alternative DM models.

\end{abstract}



\section{Introduction} \label{sec:intro}

The success of the collisionless CDM paradigm on the large-scale structure formation has made it the standard model of cosmology. The $\Lambda$CDM-based simulations \citep[e.g.,][]{springel2005_millennium, springel2008, diemand2008, stadel2009, klypin2011} have consistently shown remarkable agreement with observations on the large scale structure \citep[e.g.,][]{tegmark2006, hinshaw2013} and the model has thus far served as a strong foundation for studying many branches of astrophysical phenomena both in observational and theoretical fields, providing us deeper and enriching insights into the large-scale structure formation process.
From the other side, however, the $\Lambda$CDM model has faced outstanding challenges. 
An $N$-body $\Lambda$CDM simulation is known to produce centrally-concentrated haloes \citep{dubinski1991} in which the halo density profiles have a characteristic cuspy inner profile ($\rho \sim r^{-1}$) that is self-similar across a wide halo mass range \citep[e.g.,][]{prada2012, dutton2014}, and can be well described by a Navarro-Frenk-White profile (NFW) \citep{navarro1996, navarro1997}. The observations, on the other hand, have shown that cored profiles with shallow inner density slope $\alpha \sim 0$ in $\rho \sim r^{\alpha}$ tend to be favored in dwarf galaxies \citep{swaters2003, deblok2010, walker2011, kuziodenaray2011, oh2015}.
Related to this, the $\Lambda$CDM-predicted haloes hosting dwarf galaxies (or massive subhaloes) in the Local Group type of environment in simulations are significantly larger and more centrally-concentrated compared to observations, known as the Too-big-to-fail problem (TBTF) \citep{boylan-kolchin2011, tollerud2014, garrison-kimmel2014, papastergis2015}.
The $\Lambda$CDM model is also known to produce an excessive number of subhaloes around such environment in simulations compared to observations (missing satellites problem) \citep{klypin1999,moore1999}.

One of the favored solutions to the small-scale problems without disregarding the success of the $\Lambda$CDM on the large scale revolves around baryonic physics. 
The radiative and thermal energy output originating from the stellar feedback, including star formation and supernovae (SN) feedback could produce perturbations which disrupt and modify the gravitational potential of the central part of dwarf galaxy haloes. 
Earlier numerical studies predicted that removal of baryonic contents in the halo center by such means could lead to the formation of a cored DM profile \citep{navarro1996}.
For instance, hydrodynamical simulations that employ a bursty, stellar feedback with repeated gas outflows produced by SN explosions have shown to transform the cuspy inner profile to a shallower one, thus resolving the core-cusp problem \citep{read2005, governato2012, pontzen2012, teyssier2013, read2016, tollet2016}. The gas outflows as a result of the starbursts at higher redshift in dwarf galaxies have also been observationally implied \citep{vanderwel2011}.
Meanwhile, some studies have shown the consideration of baryon and baryonic feedback are insufficient to account for the full resolution to the problems \citep[e.g.,][]{dicintio2011, kuziodenaray2011b, parry2012}. 

The baryonic processes could also be a solution to the missing satellites problem. 
It has been proposed that in combination with stellar feedback and heating from reionization and UV background could alleviate the problems \citep[e.g.,][]{simpson2013,Onorbe2015}, while some studies show such effects are insufficient \citep{papastergis2016}.
In particular, an important role played by tidal disruption or ram pressure stripping in addition to the stellar feedback by means of removing baryonic matter from dwarf galaxies has been proposed as a solution to reconcile the problem \citep{brooks2013, sawala2013, arraki2014, brooks2014, sawala2016, wetzel2016, sawala2017}. In the meantime, \citet{trujillo2016} recently showed even with an assumption of maximal feedback effect the discrepancy in the abundance of the satellite galaxies cannot be fully reconciled. 
Taking into account baryonic processes could also be the key to solve the TBTF problem \citep[e.g.,][]{madau2014, brook2015}; but see \citet{garrison-kimmel2013, papastergis2015}.
Aside from baryons, the DM physics itself might provide an alternative solution to the small-scale problems. As pointed out by \citet{garrison-kimmel2013}, simultaneously resolving the intertwined small scale problems by baryonic processes alone still poses a challenge to the $\Lambda$CDM paradigm. A plethora of DM models have therefore been proposed without necessitating an extensive modification to the conventional $\Lambda$CDM model. Particularly interesting is the SIDM model which allows elastic scattering between DM in non-relativistic regime \citep{spergel2000}. Studies have shown that the inclusion of self-interactions of DM particles induces the creation of cores in the density profile of low-mass haloes, resolving the core-cusp problem with or without the need for baryonic processes \citep[e.g.,][]{dave2001,loeb2011, rocha2013, vogelsberger2014b, fry2015, elbert2015}.
Recent work by \citet{kamada2017} showed analytically that the observed diversity of the rotation curves from low-mass to spiral galaxies can also be addressed by the SIDM scheme.
%


The $N$-component DM model ($N$cDM) with both elastic and inelastic interactions in the dark sector is a very promising extension of the \lcdm model. The model was first proposed as a self-interacting flavor-mixed DM (fmDM) \citep{medvedev2000, medvedev2001a, medvedev2001b, medvedev2001c} in the context of dark matter halo evolution as a way to resolve the substructure problem. The inelastic DM (iDM) and exothermic DM (exDM) models were introduced in the context of the direct detection DM experiments \citep{iDM, graham2010, mccullough2013}. The excited DM (eDM) was proposed in the context of 511~keV signal in the Galaxy \citep{eDM}. Despite differences in physics of interactions in the dark sector and different evolution in the early universe, these models share much in common. They all postulate (i) the existence of more than one species, either  different "mass eigenstates'' (in fmDM) or ``excited and ground states'' (in iDM, eDM, exDM), (ii) the sufficiently large DM-DM cross sections while matter-DM interactions are of much smaller strength, (iii) the possibility of inter-conversion of the `species' in inelastic interactions which can release/absorb energy of $\Delta m c^2$ (in fmDM) or $\Delta E_i$ (in iDM, eDM, exDM). These models have nearly identical implementation in cosmological $N$-body codes, e.g., in GADGET \citep{medvedev2014, todoroki2019a} and AREPO \citep{vogelsberger2016, vogelsberger2019}.

The 2cDM model is the simplest realization of the $N$cDM. It is particularly interesting because it can resolve all the problems simultaneously, yet it does not violate all known constraints \citep{medvedev2010, medvedev2010b, medvedev2014theo, medvedev2014}. To our knowledge, 2cDM is the only model that (i) reproduces observational data, (ii) does not contradict available observational constraints and (iii) successfully and naturally evades the early universe constraint \citep{medvedev2014theo}, i.e., the Boltzmann suppression of the abundance of `excited' states after freeze-out. 

The 2cDM model is characterized by the elastic (scattering) and inelastic (conversion) cross sections, $\sigma_s(v)$ and $\sigma_c(v)$, which can be velocity-dependent, and the energy difference, $\Delta E_i$ or $\Delta m c^2$, between the two species. Numerical DM-only simulations demonstrate that $\Delta m \ll m$ (or $\Delta E_i\ll mc^2$), that is $m_1\approx m_2$ in order not to modify the large-scale structure formation \citep{medvedev2014}. It also appears that cosmological simulations can constrain the normalized values only: $\sigma_s(v)/m$,  $\sigma_c(v)/m$ and $\Delta m/m$. It is also convenient to introduce a characteristic velocity $V_k=c\sqrt{2\Delta m/m}$; we will use $V_k$ along with $\Delta m/m$. The 2cDM model's detailed theoretical foundations are described in \cite{medvedev2010, medvedev2010b, medvedev2014theo}. Note also that SIDM is automatically included in $N$cDM and corresponds to $\sigma_c(v)\equiv0$ and $N=1$.

In \citet{todoroki2019a, todoroki2019b} (correspondingly, Paper I and II), we introduced a simplistic 2cDM model, which incorporates two physical processes to the CDM paradigm: (i) the hard-sphere elastic scattering and (ii) inelastic mass conversion between two DM-species, labeled as \emph{heavy} and \emph{light} \citep{medvedev2000, medvedev2001a, medvedev2001b, medvedev2001c, medvedev2010, medvedev2014theo, medvedev2014}. 
In the model the DM cross section is generally assumed to be velocity-dependent, which arises from the quantum mechanical formalism. Such cross section's velocity-dependence has been implied as a viable possibility in simulations \citep{colin2002, vogelsberger2012, zavala2013, kaplinghat2016}.
Similarly, the 2cDM model assumes the velocity-dependent cross section for the two separate physical processes (i) $\&$ (ii) mentioned above as
\begin{equation} \label{eq:veldep}
 \sigma (v) =     \left\{ \begin{array}{ll}
        \sigma_{0} (v/v_{0})^{a_{s}} & \mbox{for scattering,} \\ 
        \sigma_{0} (p_f/p_i) (v/v_{0})^{a_{c}} & \mbox{for conversion,} 
                \end{array}\right.
\end{equation}
where $a_{s}$ and $a_{c}$ are the power-law indices of the elastic scattering and the inelastic mass conversion processes, respectively, $v_{0} = 100$ km s$^{-1}$ is the velocity normalization, and the coefficient $\sigma_{0}$ is parametrized by expressing it in terms of the cross section per unit mass, $\sigma_{0}/m$, in \units. The ($p_{f} / p_{i}$) prefactor, or $\sigma$-prefactor (or $\sigma$-prefactor), which is the ratio of the initial to the final momenta of the interacting particle, arises for the mass conversion to take into account the quantum mechanical detailed balance in the forward and reverse interaction probabilities. This prefactor explicitly appears in all cases, except for ($a_{s}, a_{c}$) = ($-2, -2$). 

Following \citet{medvedev2014}, we use the kick velocity parameter of $V_{k} = c \sqrt{2 \Delta m /m} \sim 100 \ {\rm km \ s^{-1}}$. This kick velocity depicts the boosted velocity of the \emph{light} particle that was converted from the \emph{heavy} partner after the mass conversion takes place. That is, with the mass degeneracy, we have a non-relativistic kick velocity, whereas a relativistic kick velocity is in principle possible if the difference of the two masses are assumed large. 

In Paper I $\&$ II, the 2cDM model was tested on MW-like haloes in $N$-body cosmological numerical simulations and a set of the model parameters were explored. 
We showed that the 2cDM effectively resolves the small-scale problems, namely the (i) missing satellites, (ii) too-big-to-fail, and (iii) core-cusp problems with most of the available parameters. 
To address these problems and constrain the model parameter space, the internal structure of the DM haloes and the abundance of the subhaloes were examined by looking at the halo density profiles and maximum circular velocity functions (or velocity functions for simplicity). Comparing with observations, we found that cases with the power-law indices of the velocity-dependent cross section of ($a_{s}, a_{c}$) with $a = -2, -1,$ or $0$, the self-interacting DM cross section per unit mass of $0.01 \lesssim \sigma_{0} / m$ [\units] $\lesssim 1$ and $V_{k} \sim 100$ km s$^{-1}$ can effectively solve the small-scale problems, while $V_{k} \sim 10 - 20$ km s$^{-1}$ fails to do so when the model is tested on an environment similar to the Local Group. 

Note however that these work did not consider the effect of baryonic feedback and the gas dynamics, which are non-negligible and important especially in the formation and evolution process of the MW-type halo given the large relative abundance of the luminous mass (i.e., lower mass-to-light ratio). Despite the lack of statistical samples, the parameter space for the 2cDM model was extensively explored, and their studies comprise strong implications that self-interacting, multicomponent DM model is a possibility without spoiling the success of $\Lambda$CDM on large scale.


In this Paper III, our objective is to further extend the studies presented in Paper I $\&$ II to investigate the effect of 2cDM physics to the dwarf and galaxy cluster-sized haloes.
That is, it is important to investigate whether the 2cDM model can still solve the small-scale problems across many orders of magnitude in halo mass and further deduce a tighter constraint on the set of parameters that can be compatible with observations.
For the "dwarf" simulations, we focus on the internal structure of the DM haloes by examining the DM density profiles.
For the galaxy cluster (GC) simulations, we examine a sample of GC haloes and study their internal structure by looking at both the density profiles and the fitting parameters. For this, we only focus on a particular set of parameters that are not ruled out by the dwarf simulations and the MW-sized simulations. 

The paper is structured as follows. In Section \ref{sec:simulations}, we describe the simulation setup for both dwarfs and GCs. In Section \ref{sec:dwarfs} we examine the inner structure of the DM haloes and explore the fitting parameters in comparison with observations. Further, we also present a quantitative measure on the DM velocity distributions on the radial range and the mass loss due to the inelastic mass conversion of the 2cDM model. Section \ref{sec:GCs} is dedicated to the GC simulations where we examine the DM halo density profiles and the fitting parameters to see whether the results of the 2cDM model meets the observational expectations. In Section \ref{sec:CN}, we summarize our findings and provide future prospects on the multicomponent DM model and the constrains on its parameter space.


\section{Simulations} \label{sec:simulations}

As discussed in Paper I $\&$ II, we used the same set of numerical techniques by implementing the 2cDM model in the TreePM/SPH code GADGET-3 \citep{springel2005, springel2008} on $N$-body cosmological simulations. In this work, we used two initial conditions for the Dwarf and GC simulations. The cosmological parameters were chosen to be consistent with \citet{planck2015}, where $\Omega_{m} = 0.31$, $\Omega_{\Lambda} = 0.69$, $\Omega_{b} = 0.048$, $\sigma_{8} = 0.83$, $n_{s} = 0.97$, and the normalized Hubble constant $h = H_{0} / (100 {\rm \ km \ s^{-1} Mpc^{-1}}) = 0.67$. 
All simulations start at the initial redshift of $z = 99$ and run down to the current time of $z = 0$.
For identifying the haloes and extracting their halo properties, we used the Amiga Halo Finder (AHF) \citep{knollmann2009}.


\begin{table} 
\centering
\tabcolsep=0.4cm
\begin{tabular}{ccc}
\hline\hline
  & Dwarf & GC   \\
\hline
M$_{vir}$  [M$_{\odot}$]  & $\sim$$10^7 - 10^8$ & $10^{13} - 10^{14}$ \\[0.5 ex]
$N$ of halo sample & 5 & 18-21 \\[0.5 ex]
Box size [$h^{-1}$Mpc] & 0.3 & 50    \\[0.5 ex]
$N_{\rm tot}$ & 224$^3$  & 384$^3$ \\[0.5 ex]      
$\epsilon$ [$h^{-1}$kpc] & 0.046  & 4.5 \\[0.5 ex]              
$m_{DM} $  [M$_{\odot}$]  & 309 &  2.8$\times 10^{8}$  \\
\hline\hline
\end{tabular}
\caption[]{ Summary of the simulations used in this work. M$_{vir}$ is the range of virial mass of the haloes. The box size refers to the side length of the periodic cube, $N_{\rm tot}$ is the total number of DM particles in the simulation box, $\epsilon$ is the Plummer-equivalent gravitational softening length, and $m_{DM}$ is the DM mass per simulation particle.}
\label{table:sim}
\end{table}

In our simulations, each simulation particle is a macroscopic representation of an ensemble of DM particles. We call this ensemble a simulation particle or simply 'particle'. Each of this particle is given a fixed mass, which is primarily determined by the simulation setup, such as the cosmological parameters,  the total number of simulation particles, and the size of the simulation box used.

For the purpose of studying the internal structure of the 2cDM haloes and exploring the parameter space, we use a set of high-resolution simulations. Table~\ref{table:sim} summarizes the basic parameters used for the two cases: the dwarfs and GCs. 
For dwarfs, we used a rather small cubic box of 300 $h^{-1}$kpc per side length with $224^{3}$ particles. In such a small simulation box, the size of largest halo that can be produced is limited to the order of $10^8$ M$_{\odot}$, and the strong environmental effects, such as the tidal stripping that could originate from the host halo, are therefore absent. Thus, the setup is rather close to an isolated dwarf halo and strictly speaking it is not cosmological. The largest halo on the order of $10^{8}$ M$_{\odot}$ contains more than a million simulation particles with a single DM mass of 309 M$_{\odot}$. The force resolution is set to 46 $h^{-1}$pc, which is small enough for our purposes to study the internal structure of the five largest haloes over the range of M$_{\rm vir} \sim 10^7  - 10^8$ M$_{\odot}$.

For GCs our sample contains haloes of the order $10^{13} - 10^{14}$ M$_{\odot}$. 
The simulation box size is 50$h^{-1}$Mpc for the side length and the total number of particles is 384$^{3}$. The force resolution is about two orders of magnitude larger than the dwarf simulation (i.e., 4.5 $h^{-1}$kpc), but it provides enough accuracy in the inner radial profiles to ascertain whether the 2cDM is capable of creating shallower inner slope as it is indicated by observations for galaxy clusters. The total number of haloes studied ranges from 18 to 21, depending on the choice of the parameters in which some produced a few outliers that are mostly attributed to numerical artifacts.

To ease the comparison among the models based on the different set of parameters, we use the same initial condition that was used for all cases tested on each setup for dwarf and GC simulations. 
The cosmological parameters are also unchanged for all cases in order to see the direct effect of each set of 2cDM parameter on the halo properties.



\section{Dwarf Haloes}  \label{sec:dwarfs}

It has been observationally shown that low-mass galactic haloes, including Low Surface Brightness (LSB) galaxies and dwarf spheroidals (dSphs), tend to have shallower rotation curves in the inner radial profile \citep{deblok2008, deblok2010, oh2011, oh2015}, which concurrently implies cored halo density profiles, as opposed to the cuspy profiles predicted by $N$-body numerical simulations of a $\Lambda$CDM cosmology \citep{flores1994}. 
There is a mixed conclusion in literature that some claim a cascade of early supernovae feedback can transform a cuspy inner profile to cored one in dwarf galaxies \citep[e.g.,][]{navarro1996b}, while others argue that in such DM-dominated systems star-formation-induced energetic supernovae are inefficient to achieve such transformation based on observational constraints \citep[e.g.,][]{kuziodenaray2011}.
Here we study whether the 2cDM physics alone could sufficiently explain the formation of cored density profiles in dwarf haloes without considering the presence and the effects of baryonic physics. 

To begin, we first present the halo density profiles and examine the internal structure based solely on the DM mass distribution. Subsequently, the parameters are constrained by applying the fit to the profiles and comparing it with observations. We then study the direct effects of the elastic scattering and mass conversion (or "quantum evaporation" effects) of 2cDM in the DM velocity profiles as well as the DM velocity distribution function within a halo. The phase-space diagram is also shown to check the effects.
Some of the selected set of parameters are further studied to see the effects of 2cDM on the anisotropy radial profiles and the halo spin parameter. 
Finally, in comparison with the CDM counterpart, we quantify the fraction of halo mass that can be lost or evaporated by the 2cDM physics.

A summary of the set of parameters explored in our dwarf simulations are the following:
(i) $\sigma_{0} / m$ = 0.001, 0.01, 0.1, and 1 \units, (ii) ($a_{s}, a_{c}$) = ($X, Y$) where $X, Y = -2, -1, 0$, which gives 9 cases in combination. 
The kick velocity $V_{k} = 100$ km s$^{-1}$ is used throughout this work as the fiducial value, which corresponds to the mass degeneracy of $\Delta m / m \sim 10^{-8}$ (see Section \ref{sec:intro}). 
Most of these parameters are chosen in accord with the results from Paper I $\&$ II.

\subsection{Density profiles}

\begin{figure}
  \centering
  \includegraphics[scale = 0.47]{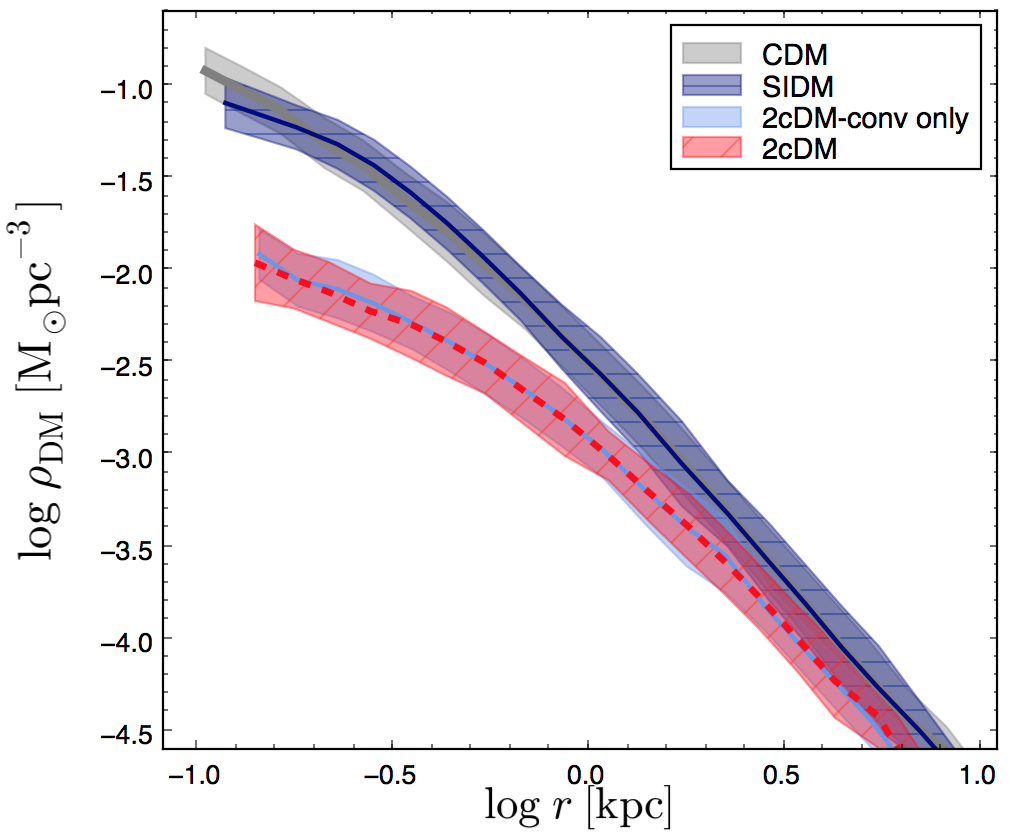}  
  \caption[]{\label{fig:dwarf_profile-00-sig01} %
The mean DM density profiles of 2cDM for dwarf haloes compared with selected models. The solid or dashed curves are the mean and the shade is 1 $\sigma$ standard deviation among the sample of five most-resolved haloes.
The inner-most radial range where numerical convergence fails based on two-body collision criteria is not shown.
  }
\end{figure}

Figure \ref{fig:dwarf_profile-00-sig01} shows the mean DM halo density profiles for the selected models to highlight the effect of 2cDM.
We chose $(0,0)$ (i.e., no velocity dependence) with $\sigma_{0} / m = 0.1$ \units for all the cases shown: SIDM (= elastic scattering only), 2cDM$^{\rm conv}_{\rm only}$, and 2cDM. The CDM is also shown for a comparison. The sample consists of the largest five haloes in the simulation box.
It clearly shows the mass conversion is the key physical process that successfully reduces the inner-most density, as 2cDM$^{\rm conv}_{\rm only}$ and the full 2cDM (both mass conversion and elastic scattering enabled) closely follow each other's trend. 
Their profiles start to deviate from the CDM and SIDM at $\sim 1$ kpc with the chosen set of parameters, generally conforming to the observed range.

\begin{figure*}
  \centering
  \includegraphics[scale = 0.50]{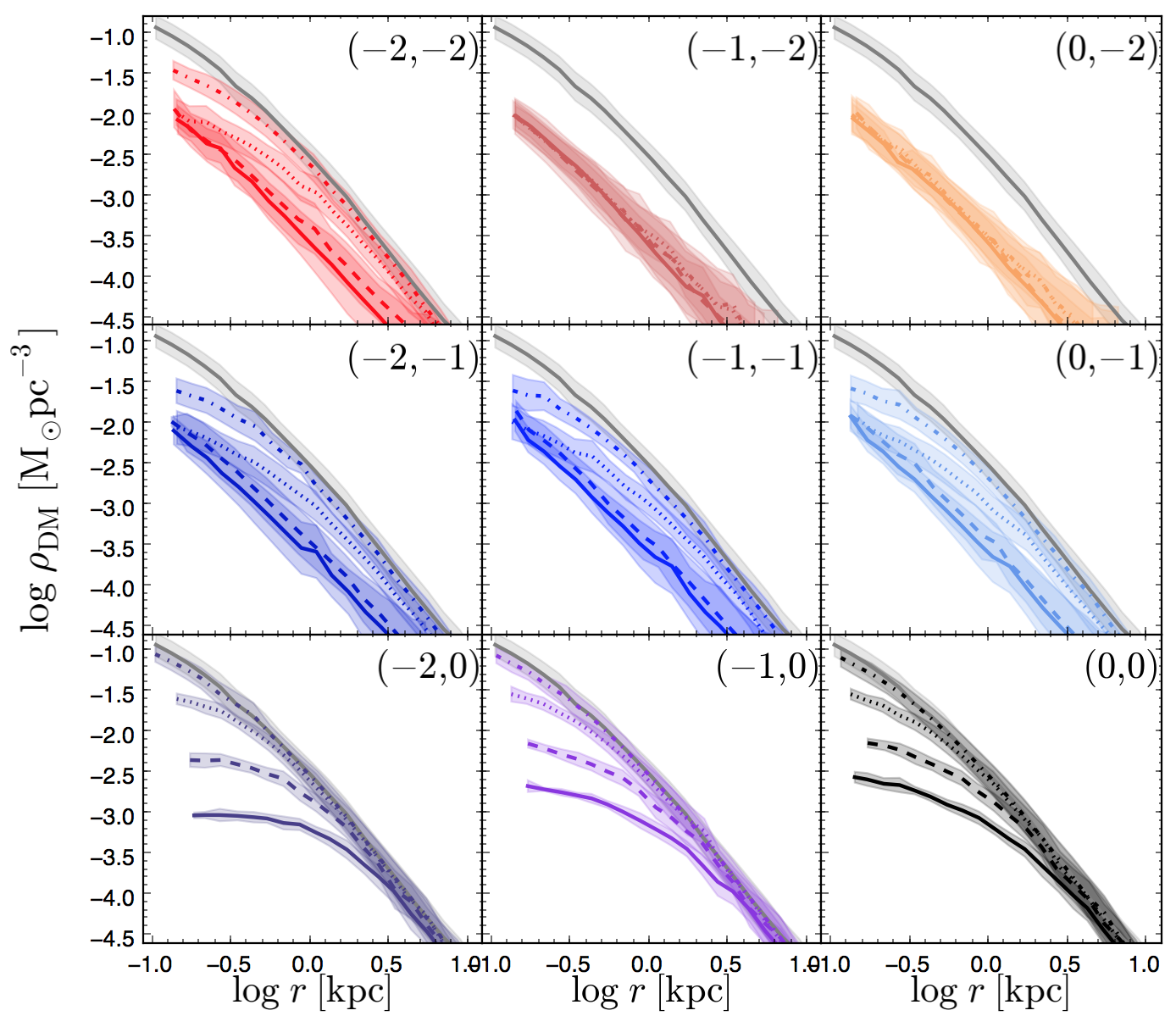}  
  \caption[]{\label{fig:dwarf_profile_all_orig4} %
The mean with a 1 $\sigma$ standard deviation of DM halo density profiles of the 2cDM models compared with the CDM model (gray solid curve). The number of halo samples used was 5 (Table \ref{table:sim}). The dash-dot, dotted, dashed and solid curves represent $\sigma_{0}/m = 0.001$, 0.01, 0.1 and 1 \units, respectively. 
  }
\end{figure*}

To explore the parameter space further, Figure \ref{fig:dwarf_profile_all_orig4} shows the compilation of profiles for all the other cases of 2cDM.
We immediately see the prominent impact of $a_{c}$ on the formation of a cored density profile by simply comparing the rows, whereas that of $a_{s}$ is minimal by comparing across the columns. In other words, the shape of the profile is predominantly determined by the strength of mass conversion rather than elastic scattering. This is particularly true for low-mass systems such as the dwarf haloes considered here; their intrinsically small DM velocity has a significant effect on the cross section that is inversely proportional to the velocity.

The following are a list of other implications:  
(i) The shape of density profile appears nearly identical, irrespective of the value of $\sigma_{0} / m$, for the cases with the mass conversion having a strong velocity dependency of ($X, -2$) where $X = -2, -1$, or $0$. 
It is unclear what the exact physical mechanism is attributed to such similarity. The exception is ($-2,-2$), in which case the $\sigma$-prefactor becomes identity and behaves similar to the case with $(-2,-1)$. 
Meanwhile, a strong evaporation effect is clearly responsible for creating less steep logarithmic slopes over the haloes' virial range compared to the CDM. It is likely that its dominant effect is negating the $\sigma_{0} / m$-dependency in this case.
(ii) The models with $\sigma_{a_{c}} \sim 1 / v$, or ($X, -1$), show cored density profiles only for smaller $\sigma_{0} / m$ values of 0.001 and 0.01 \units. For larger $\sigma_{0} / m$ values ($\gtrsim 0.01$ \units), we are unable to tell whether the core formation is possible within the radial scale our simulations can resolve. 
(iii) For ($X$, 0) where the mass conversion cross section has no velocity-dependence $\sigma_{a_{c}} \neq \sigma_{a_{c}} (v)$, the core density ($\rho_{c}$) and core radius ($r_{c}$) are primarily determined by $\sigma_{0} / m$, producing self-similar profiles. This particular model shows a clear trend where a smaller $\rho_{c}$ (and a larger $r_{c}$) is created by a larger $\sigma_{0} / m$. 


To quantitatively compare the results with observations and constrain the parameters, we use a cored density profile model, which is a modified version of the isothermal (ISO) model.  
Following the formula introduced in Paper I $\&$ II, we fit the dwarf halo density profiles with the generalized isothermal model (gISO), which is given by
\begin{equation} \label{eq:gISO}
\rho_{\rm gISO}(r) = \rho_{c} \left[ 1 + \left(\frac{r}{r_{c}} \right)^{2} \right]^{- p / 2},
\end{equation}
where $p$ is a parameter that introduces a flexibility to the pure ISO model for the outer slope of the density profile. Note that with $p \rightarrow 2$ the model is effectively reduced to the pure ISO model.
This model inevitably gives a poorer fit to the cases with cuspy inner profiles.
For example, most of ($X, -2$) and ($X, -1$) cases clearly do not show a sign of core formation within the resolved radial scale, which corresponds roughly to $\sim$100 pc.

To ease the comparison with observations and mitigate the problem arising from the fit, we take the total halo mass within 300 pc from the halo center (M$_{300}$ = M$(r \leq 300 {\rm pc})$) instead of $\rho_{c}$, which tends to be poorly determined, especially for the cuspy cases. The advantage of M$_{300}$ over $\rho_{c}$ is that it is a parameter that can simply be determined by the number of DM particles reside in $r \leq 300$ pc independently of the fitting model used. It also allows us to quantify the effectiveness of the mass evaporation directly.  
For completeness, however, we conducted the fit on all cases with gISO over the numerically resolved radial range. 

\begin{figure*}
  \centering
  \includegraphics[scale = 0.50]{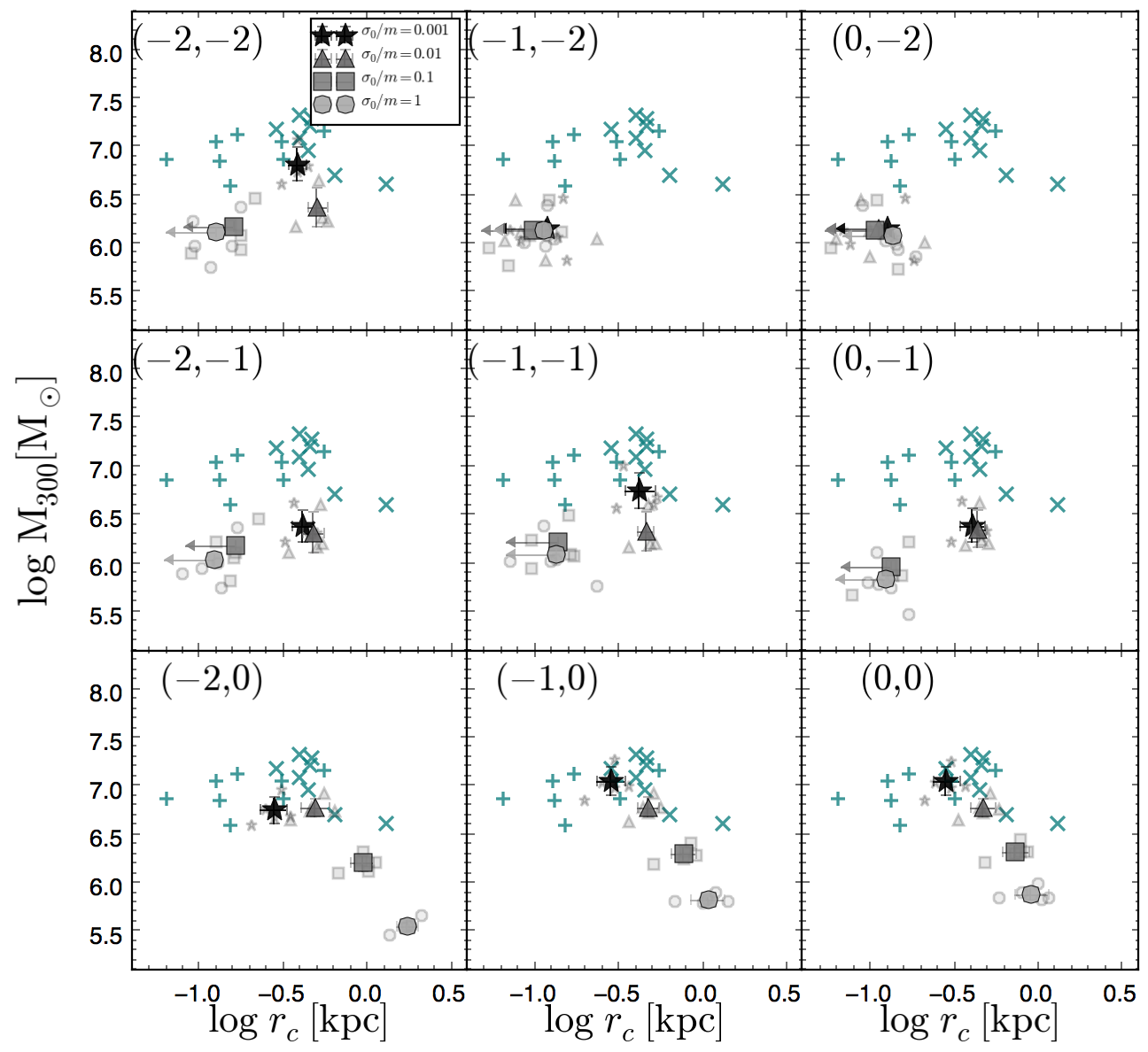}  
  \caption[]{\label{fig:dwarf_profile_all_fitparam} %
The fitting parameters, $\rho_{c}$ and $r_{c}$, of the 2cDM models compared with observations. 
The DM cross section per unit mass (in \units ) over four orders of magnitude are shown for each model with the mean and the 1$\sigma$ error bars and the five individual haloes with the faded color. For models which showed "cuspy" profiles, the upper-limits on $r_{c}$ are shown since the gISO fitting model, a cored profile, tends to overpredict it, while underpredicting $\rho_{c}$. The crosses are taken from observational data on MW dSphs \citep[and references therein]{strigari2008b, burkert2015}.
  }
\end{figure*}

Figure \ref{fig:dwarf_profile_all_fitparam} shows M$_{300}$ as a function of $r_{c}$ along with observational data from Milky Way dSphs.
For the reasons mentioned above, we take $r_{c}$ for the cuspy ones as the upper-limits.
As expected, those 'cuspy' cases consistently show smaller M$_{300}$ due to stronger evaporation effect induced by the stronger velocity dependence of the cross section with either $a_{c} = -2$ or $-1$.
The only cases that show good agreement with observations, and thus are unlikely to be ruled out are: ($-2,-2$), ($-1,-1$) and ($X,0$), where ($X=-2,-1,0$), with $\sigma_{0} / m = 0.001$ and 0.01 \units.
In the meantime, $\sigma_{0} / m = 0.1$ and 1 \units are unlikely, at least from the $N$-body simulation presented here.

%

\subsection{Velocity profiles}

\begin{figure}
  \centering
  \includegraphics[scale = 0.34]{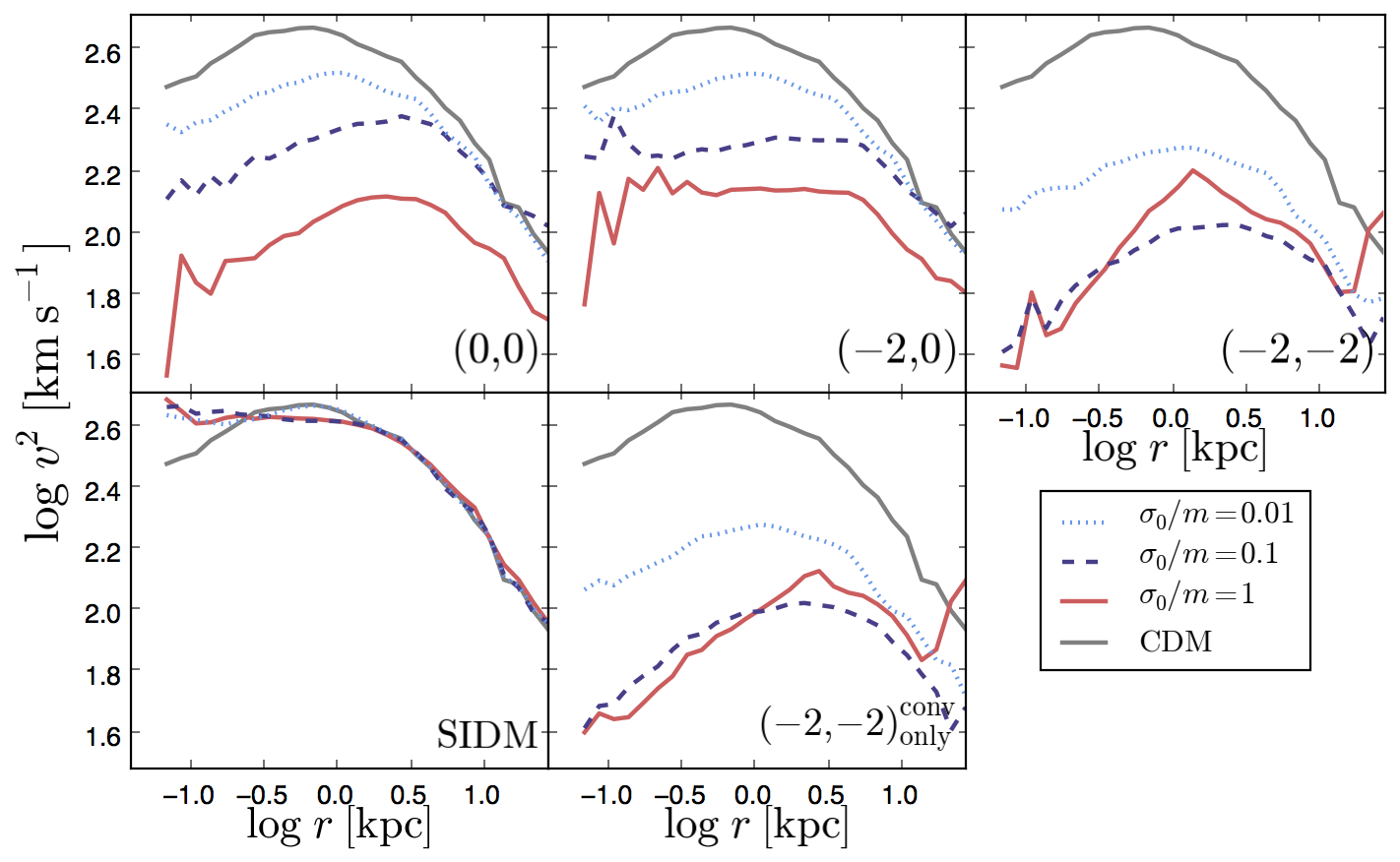}  
  \caption[]{\label{fig:v2_profiles} %
Velocity profiles of the most-resolved halo for the selected models. The velocity is the mean at each of the spherically radial bin. $\sigma_{0}/m$ is in \units.
  }
\end{figure}

By examining the density profiles, it is clear that the inelastic mass conversion process plays a significant role in characterizing the DM mass distribution within the dwarf haloes. 
As briefly mentioned in Section \ref{sec:intro}, the most important feature of the mass conversion is the non-relativistically boosted \emph{light} particles escaping the halo, altering the velocity distribution of DM.
To highlight the effect, Figure \ref{fig:v2_profiles} shows the velocity profiles, $v^{2}(r)$, of the selected cases.
Here we focus on examining the effect of the elastic scattering and the mass conversion on the shape of the profile by comparing the cases with (i) no velocity-dependence of the cross section $\sigma$ for both elastic scattering and inelastic mass conversion, i.e., ($0,0$), (ii) the velocity dependence of $\sigma$ is only applied for the elastic scattering with $a_{s} = -2$, or ($-2,0$), (iii) a case with strong velocity dependence of $\sigma$, ($-2,-2$), and its duplicate case with the elastic scattering process disabled, $(-2,-2)^{\rm conv}_{\rm only}$, and (iv) another ($-2,-2$) case with the mass conversion disabled (or equivalently, SIDM). 
A range of $\sigma_{0} / m$ values are also checked for each case. 
The profile is computed by taking the mean of $v^{2}$ of DM particles that reside within each radial bin in a spherical shell. For a clear comparison, the same most-resolved halo ($M_{vir} \sim 10^{8}$ M$_{\odot}$) was used to extract the profile for all cases.

The most prominent outcome of the evaporation seen in the velocity profile is that the high-velocity particles are being preferentially removed from the halo due to the reduction of the halo mass.
Although the inner-most part of the halo is where the DM interaction rate is the highest, the boosted DM particles in a small halo with $V_{\rm max} \lesssim 20$ km s$^{-1}$ easily escape from the halo potential after the mass conversion. 
This is clearly seen in the conversion-only case of $(-2,-2)^{\rm conv}_{\rm only}$ and ($-2,-2$) in which a significant loss of the high-velocity DM particles from the halo is obvious. 
The elastic scattering, in the meantime, causes mere exchange of the kinetic energy of the interacted particles and does not lead to any loss of halo mass (i.e., the case for SIDM). The primary effect of such exchange of energy is the increase of velocity dispersion in the inner-most part of the halo, producing a flattened profile.
This demonstrates the mass conversion has the dominant effect on determining the shape of the velocity profile, just as it is so for the mass density profile. 
 
\begin{figure}
  \centering
  \includegraphics[scale = 0.34]{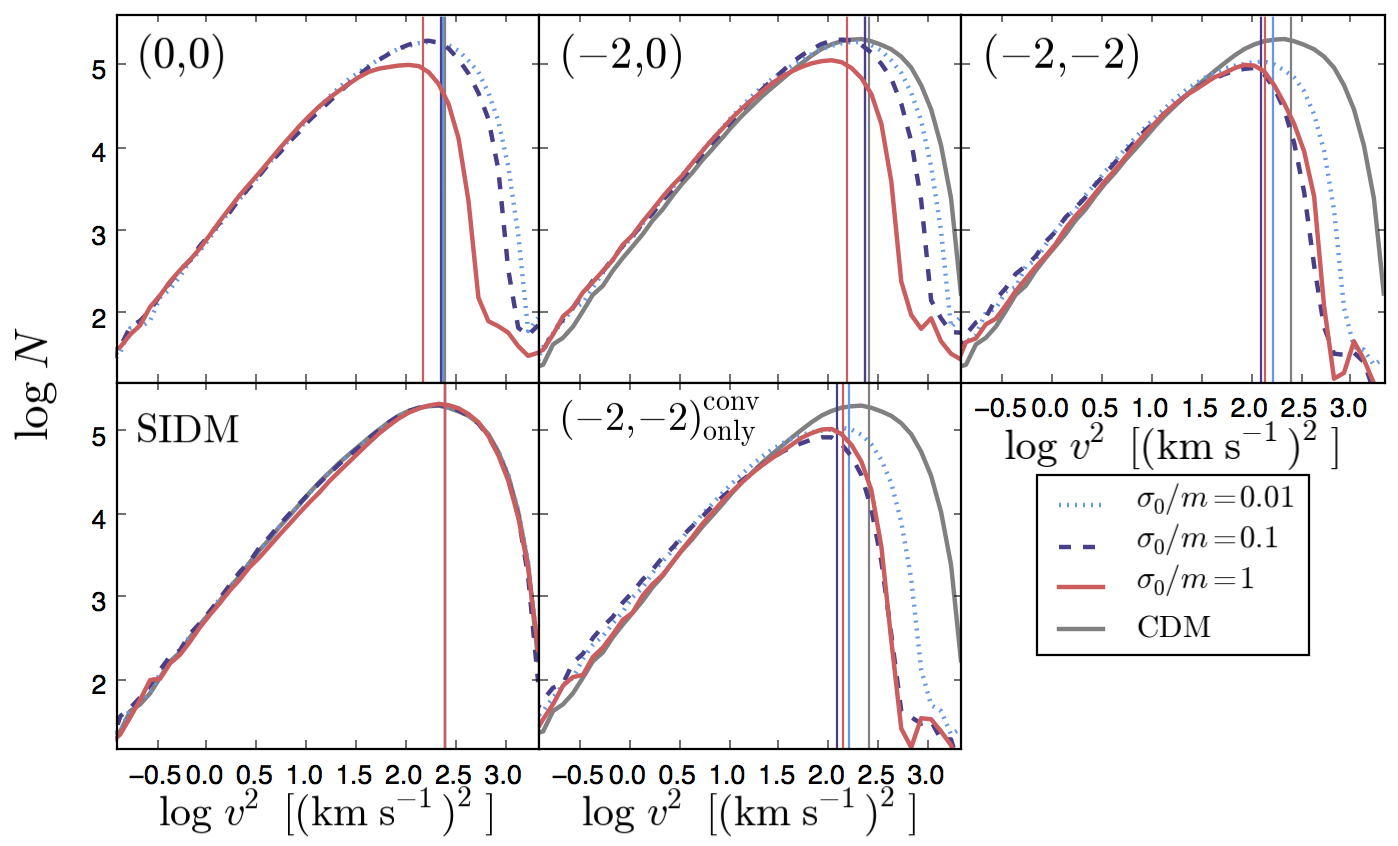}  
  \caption[]{\label{fig:N_vs_v2} %
Distribution function of $v^{2}$ for the most-resolved halo for the selected models. The vertical lines represent the escape velocity squared ($V_{esc}^2$) of the halo for each model. $\sigma_{0}/m$ is in \units.
  }
\end{figure}

To quantitatively see it from another point of view, Figure \ref{fig:N_vs_v2} shows the distribution function of $v^{2}$, the mean of the squared velocity in the spherical radial bin. The turnover occurs roughly around the escape velocity of the halo (indicated as vertical lines), and it shifts to the lower velocity side for the cases where strong evaporation is observed.
The identical distribution function of SIDM with respect to CDM signifies the elastic scattering alone does not lead to the reduction of the halo mass and has a negligible effect on the velocity distribution function within the halo.

\begin{figure*}
  \centering
  \includegraphics[scale = 0.5]{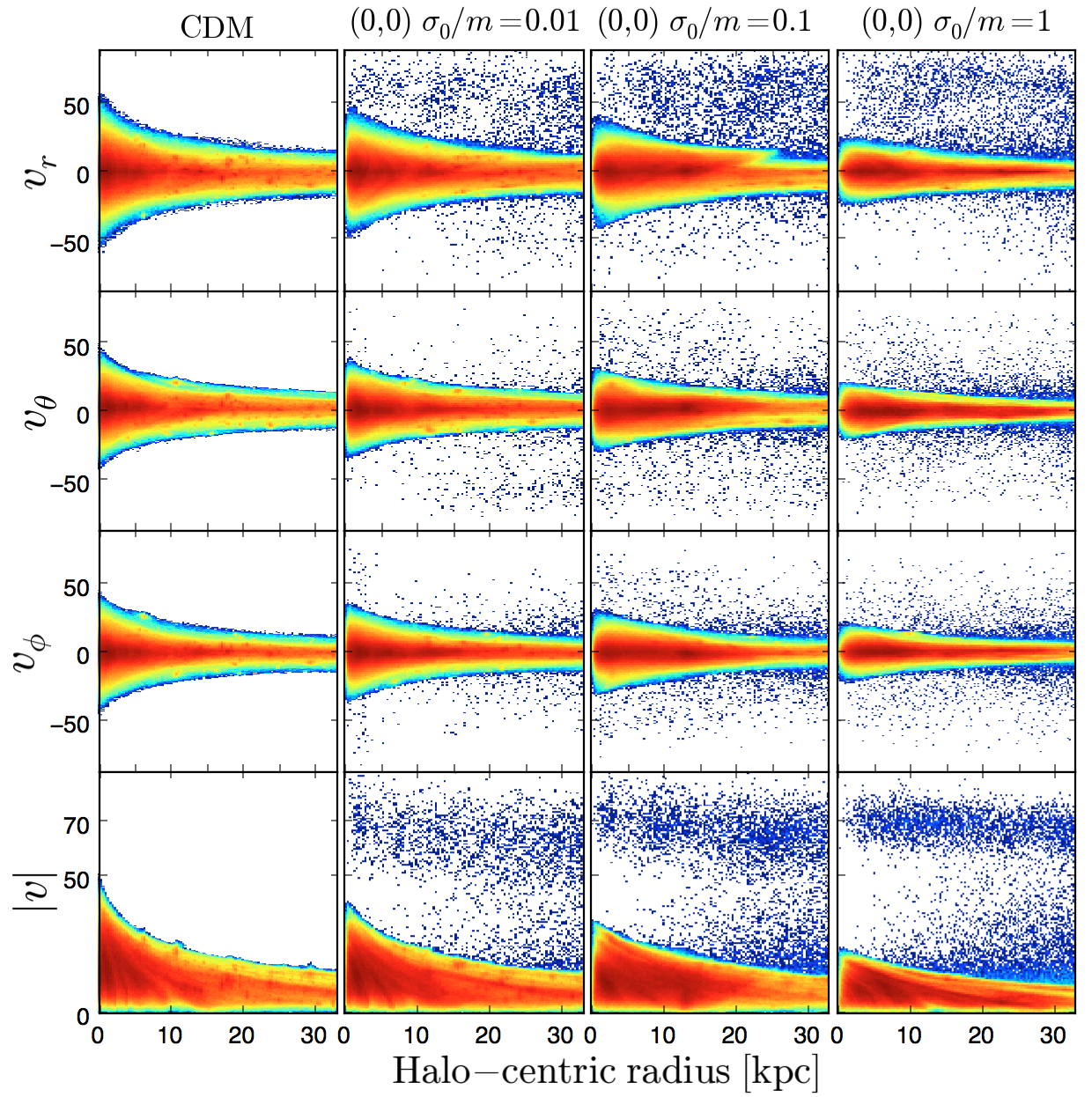}  
  \caption[]{\label{fig:phase-space} %
Phase space diagram portrayed as velocity profile of the most massive halo. To illustrate the difference, $(0,0)$ with $\sigma_{0}/m = 0.01, 0.1$ and 1 \units is displayed and compared with the CDM. Each pixel represents a simulation particle. The color scheme used here is based on the projected mass density with red being the densest and blue being the least dense.  
  }
\end{figure*}

Another effective way of examining the consequence of the mass conversion is to study the phase space.  
Figure \ref{fig:phase-space} shows the phase space diagram in the form of spherical velocity profiles of the halo. In this case each pixel represents an individual DM simulation particle. To illustrate the key point, only the ($0,0$) case is shown in comparison with CDM. 
It captures the essence of the mass conversion effect of the 2cDM model as seen in the boosted \emph{light} particles being relocated to well outside of the virial radius of the halo ($R_{vir} \lesssim 23$ kpc).
This creates a stark contrast with the CDM model shown in the left-most column. 
It is also interesting to see how the abundance of the substructures is reduced in the 2cDM model compared to the CDM. Substructures appear as high concentrations of the particles in the phase space, and such signature is smoothed out and disappear for ($0,0$) with a larger cross-section value.

\begin{figure}
  \centering
  \includegraphics[scale = 0.41]{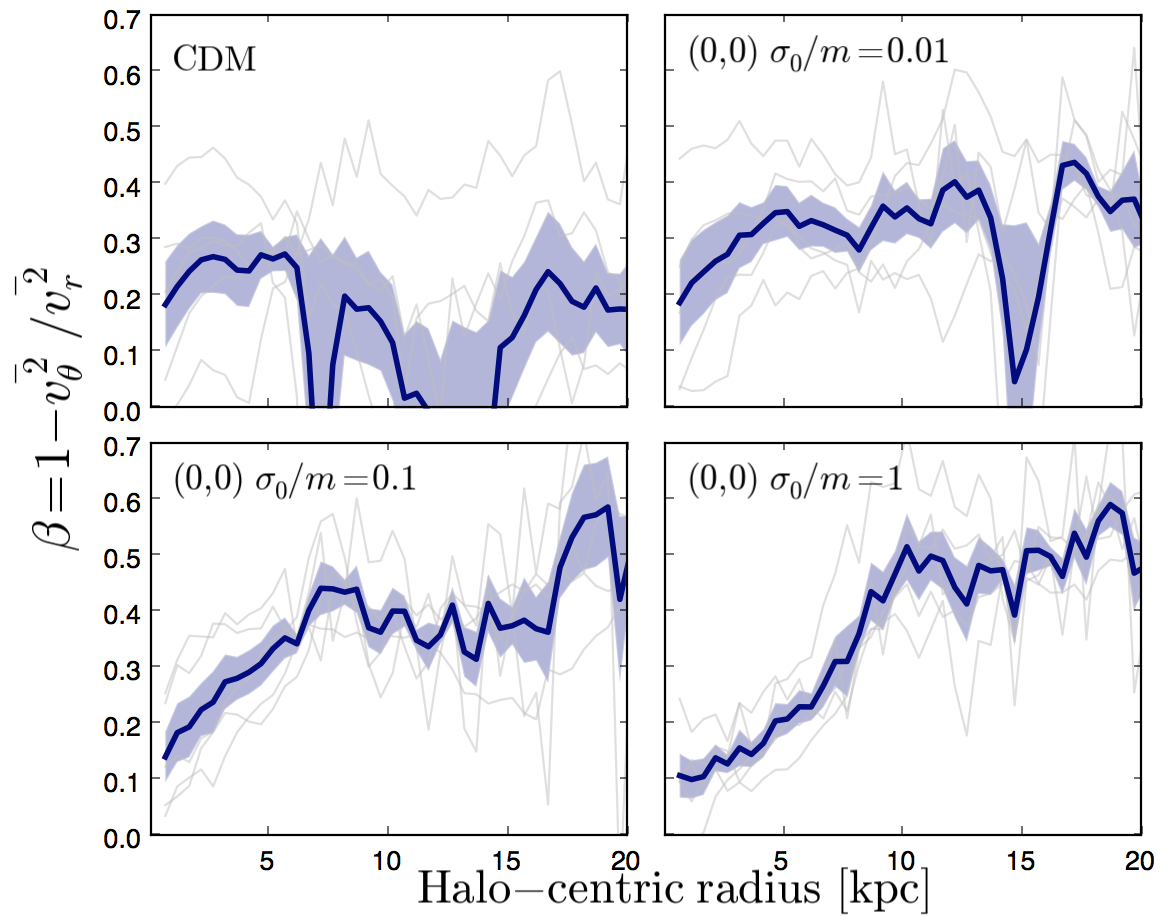}  
  \caption[]{\label{fig:beta_profile} %
Mean anisotropy profile with the standard error from the sample of five most revolved haloes. The individual halo profiles are also shown in thin gray. $\sigma_{0}/m$ is in \units.
  }
\end{figure}

One could extend the study of the halo structure based on the velocity-component by examining the anisotropy of the halo.  
The so-called anisotropy parameter is defined as $\beta \equiv 1 - \bar{v^{2}_{\theta}} / \bar{v^{2}_{r}}$ \citep{binney2008}, which describes the geometry of the internal structure of the halo in terms of the velocity in the spherical coordinates. Then the sphericity, or isotropy, corresponds to $\beta \sim 0$ with $\bar{v^{2}_{\theta}} \approx \bar{v^{2}_{r}}$, while the degree of anisotropy increases as the radial component dominates over the polar component ($\bar{v^{2}_{\theta}} \ll \bar{v^{2}_{r}}$), giving $\beta \rightarrow 1$. We evaluate the mean anisotropy profile $\beta(r)$ of the five most resolved haloes and it is presented in Figure \ref{fig:beta_profile}.

Despite the statistically poor sample, we find a clear transformation of the degree of anisotropy within a halo for the ($0,0$) cases. Notably, the rise of $v_{\theta}$ relative to $v_{r}$ (hence, declining $\beta$ value) near the halo center is consistently seen among the halo sample for the case with larger cross section of $\sigma_{0}/m = 0.1$ and 1 \units that are also producing a clear cored inner density profile. We can see that their halo structure is divided into two regimes -- one that is more isotropic ($\beta \rightarrow 0$, inner part) and the other being more anisotropic (outer part of halo). The boundary which separates these two regimes roughly corresponds to the characteristic radius in the density profile where the shape of the profile transforms from $r^{-3}$ to a shallower, cored one. 
When the cases with $\sigma_{0}/m = 0.1$ and 1 \units are compared, one can see that the size of the characteristic radius increases for the latter case, indicating the expansion of the spherical core region inside the halo due to the stronger mass evaporation effect accompanied with a larger interaction rate.

\subsection{Mass loss fraction}

\begin{figure}
  \centering
  \includegraphics[scale = 0.38]{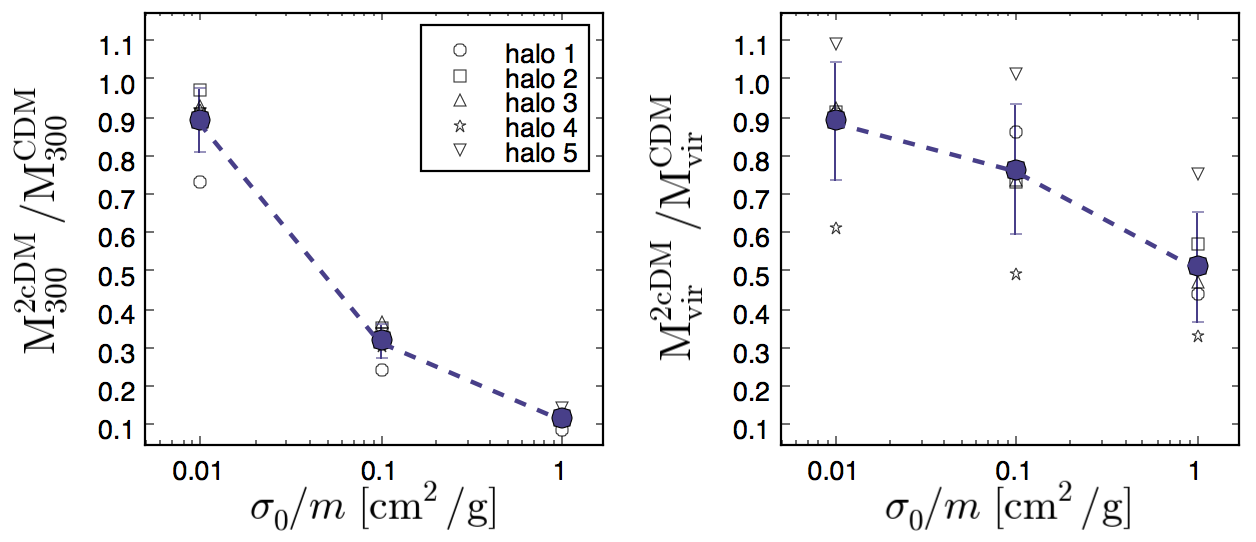}  
  \caption[]{\label{fig:M2cDM_by_Mcdm} %
Relative halo mass of 2cDM and CDM for ($0,0$) at $z = 0$. The solid points represent the mean with $1 \sigma$ error bar and open points are individual haloes with halo 1 being the largest in terms of the virial mass. 

  }
\end{figure}

We now quantify the fractional mass loss due to the 2cDM physics, namely the mass conversion, on individual halo bases. 
In Figure \ref{fig:M2cDM_by_Mcdm} we compare the ratio of 2cDM halo mass to that of the CDM counterpart as a function of the DM cross section at $z = 0$ and examine how much mass is lost from (i) the inner part of $r < 300$ pc (of which the halo mass contained within is denoted as M$_{300}$) (left panel) and from (ii) the halo virial radius of roughly $R_{vir} \sim 20$ kpc or less for our halo sample (right panel). We chose the (0,0) model, where the cross section has no dependency on the velocity for elastic scattering, while mass conversion process has a dependency of $1/v$ that arises from the $\sigma$-prefactor.

By comparing the left and right panels, it is immediately clear that the mass loss is more substantial in the inner part than the virial range. 
This in turn means not all boosted particles fully escape the halo and some fraction of them remain bounded to the potential after mass conversion. For example, the case with $\sigma_{0}/m =1$ \units shows the mass reduction achieved in the inner part is $\sim 90\%$, whereas that of over the virial range is only $\sim 50\%$ (on average) relative to the CDM halo mass. 
For the case with $\sigma_{0} / m = 0.01$ \units, the mass reduction is kept minimal ($\sim 10\%$) in both regimes.
In either case, our results indicate that small haloes with M$_{\rm vir} \sim 10^7 - 10^{8}$M$_{\odot}$ are not completely blown away by losing all of its mass even with a case with a strong interaction rate provided by a large $\sigma_{0} / m$ value. 

Note that a more accurate representation is $M_{300}$ than $M_{vir}$ since the boundary of halo may not be well-defined, resulting in a larger scatter among the ${\rm M^{2cDM}_{vir} / M^{CDM}_{vir}}$ sample compared to that of ${\rm M^{2cDM}_{300} / M^{CDM}_{300}}$.
We emphasize that the mass loss fraction presented here is predominantly due to the mass conversion and is independent of environmental effects, such as tidal stripping, since the small box size we use does not contain any other large haloes.
Interestingly, it appears that the fractional mass loss does not seem to strongly depend on the size of halo with this size; that is, the largest halo (open circle) is not necessarily the one that shows the strongest reduction of mass compared to the other smaller haloes within our sample (${\rm M_{vir}}\sim10^{7} - 10^8 M_{\odot}$). 
This implies the presence of non-negligible statistical fluctuations among the sample, hence, for a more accurate analysis a larger sample is clearly desired.

\section{Galaxy cluster haloes} \label{sec:GCs}

As the largest gravitationally bound objects found in the Universe, GCs offer a crucial venue to explore and study the 2cDM model on the high-mass end of the halo mass function. 
Similar to dwarf galaxies, GCs are DM-dominated with high mass-to-light ratios, which makes them best suited for studying the role of DM played on cosmological scales. 
According to the bottom-up scenario, small structures form in the early Universe continuously accrete mass and experience mergers by gravity over the comic time scale to form GCs at later times. 
With such a long time scale evolutionary process, DM haloes grow in size and can cover many decades of mass range, which places dwarf galaxy haloes and GCs at the both ends of the extreme in halo mass.
In this section, we study whether the 2cDM model is capable of reproducing agreement with observations by further constraining the model parameter on GC haloes. 

({\it Simulations}) 
To achieve our goal, we test the 2cDM model on GCs by examining the internal structure of the halo and compare the results with observationally available data. We chose a set of simulations with the total number of simulation particles of $384^{3}$ with a cubic side length of 50 $h^{-1}$Mpc. The force resolution is set to 4.5 $h^{-1}$kpc, which is small enough to allow us to discern whether the given set of parameters can be ruled out. That is, some observational studies have shown that a typical GC core size can be $r \lesssim 50 - 100 $ kpc \citep{allen2011}, and hence if the 2cDM GC core size either exceeds 100 kpc or is much smaller than 50 kpc, we can safely rule out the particular set of model (although this requires caution since it is subject to the baryonic effects as well, which is discussed in the Section \ref{sec:discussion}). 
We explored some of the most promising cases that have survived the parameter studies on the MW-sized in Paper I $\&$ II and dwarf-sized haloes (Section \ref{sec:dwarfs}). 

({\it Halo sample})
The simulation box size is large enough, for our purposes, to have a GC sample of $\sim$20 in the range log$M_{200} / M_{\odot} \sim 13.5 - 14.5$ with the mean of $13.9 \pm 0.3$. 
We study the sample taken at $z = 0.25$, which corresponds to the redshift of some of the observed GCs we use for comparison. 
The haloes were selected based on the total number of particles contained within the virial halo radius, $N_{\rm halo} (< R_{\rm vir}) > 100,000$. We set $R_{vir} = R_{200}$, that is, the GC halo virial quantities are estimated based on the density contrast that exceeds 200 times of the critical density of the Universe. This yields about 20 haloes on average in our simulations over the mass range mentioned above. Our sample consists of both dynamically relaxed and unrelaxed haloes. 


\subsection{Density profiles}

\begin{figure}
  \centering
  \includegraphics[scale = 0.40]{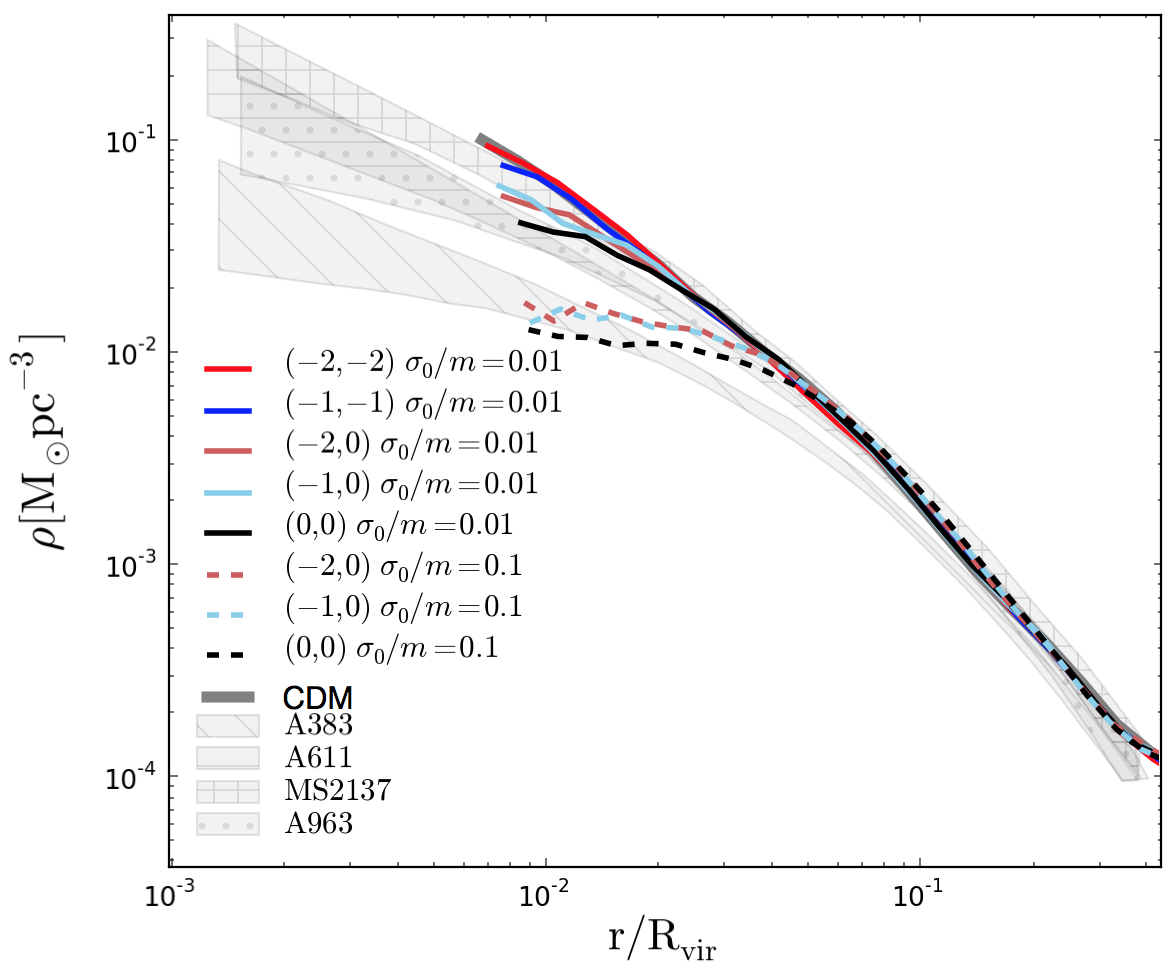}  
  \caption[]{\label{fig:GC-profiles} %
The mean DM density profiles of selected 2cDM models for GC haloes compared with observations (from \citet{newman2013b}).
The solid or dashed curves are the mean. The standard deviation is not shown in order not to lose the clarity. 
The inner-most radial range where numerical convergence fails based on two-body collision criteria is not shown. 
$\sigma_{0}/m$ is in \units.
  }
\end{figure}

\begin{figure*}
  \centering
  \includegraphics[scale = 0.5]{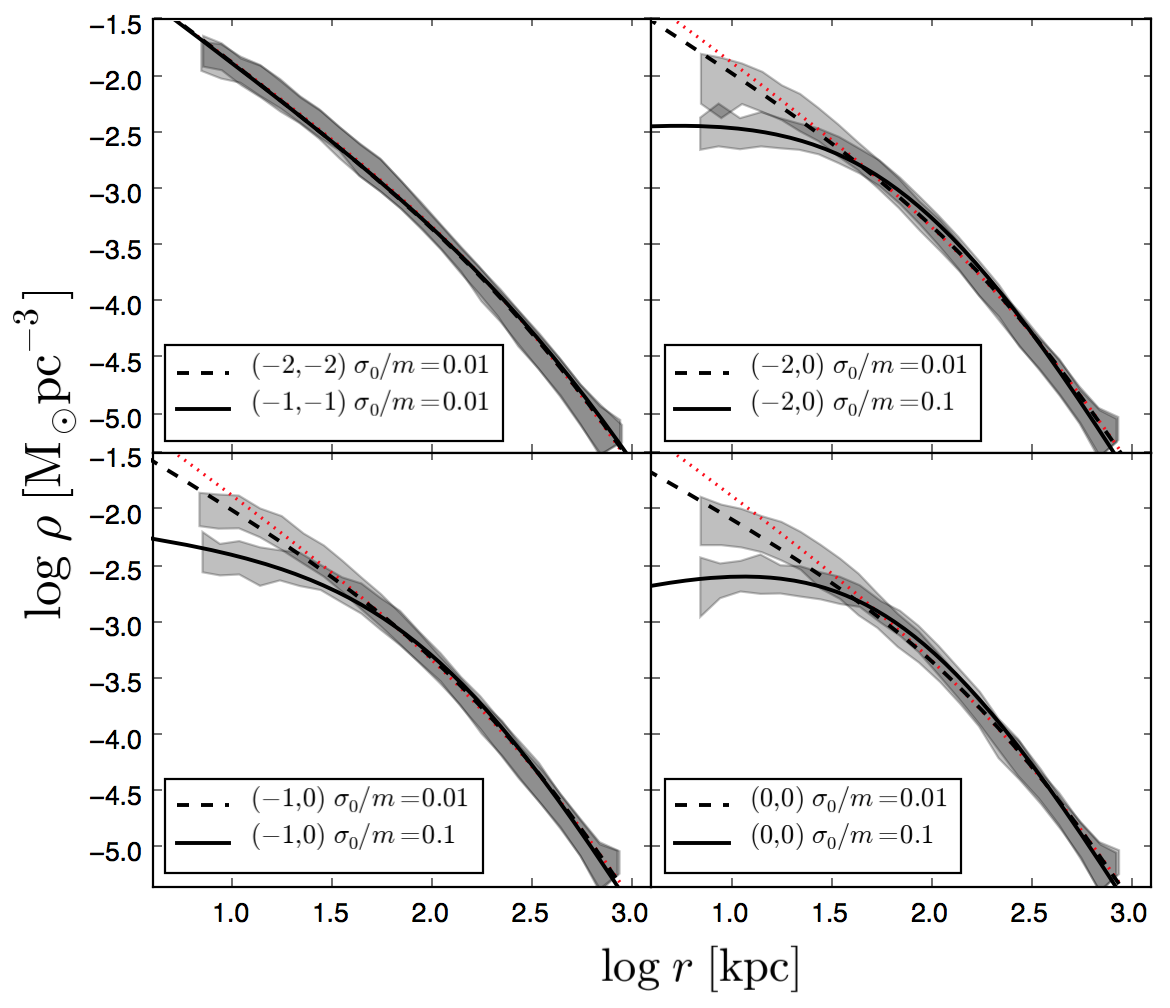}  
  \caption[]{\label{fig:GCprofile} %
Mean radial density profiles with 1$\sigma$ spread for the sample of GC haloes in comparison with CDM (red dotted). The gNFW fit are shown in solid and dashed curves. $\sigma_{0}/m$ is in \units.
  }
\end{figure*}

Observationally, the mass distribution of GCs are probed by gravitational lensing, X-ray emission, and optical observables.  
In particular, taking the advantage of the deep gravitational potentials produced by GCs, gravitational lensing technique provides a robust way of probing the mass distribution of GCs regardless of whether it is luminous or dark matter. It also has an advantage of not necessitating the assumption of the hydrostatic equilibrium unlike in X-ray observations. Based on Einstein's theory of general relativity, a presence of mass or deep gravitational potential, such as in GCs, creates curvature in its surrounding space-time and deflects the path of light rays, resulting in the distortion patterns seen in the image of the distant background galaxies. 
Mass distribution can be mapped by measuring such distortion that appears as giant arcs centered around the gravitational potential (strong lensing), and by systematically studying a weaker and more coherent distortion patterns on the image of background galaxies (weak lensing) \citep[e.g.][]{bartelmann2001}. 
A disadvantage of the lensing method, especially relevant to the strong lensing, is that it is sensitive to the mass projection bias due to the triaxial halo shape \citep[e.g.][]{white2002, torri2004, gavazzi2005, hennawi2007}. The CLASH cluster survey \citep{postman2012} used a selection criteria of clusters based on X-ray morphology specifically for avoiding such disadvantage.

The conventional $\Lambda$CDM in $N$-body simulations has shown to reproduce a cuspy density profile for GC haloes which is well described by an NFW profile. This raises a possible tension with some observational studies that have shown a flat or mildly cuspy inner density profiles in the observed GC haloes \citep{ettori2002, sand2004, newman2011,newman2013b}, whereas other studies have shown that there is no such tension \citep{schmidt2007}.
Here we check whether the 2cDM model is capable of producing a density profile that is consistent with both of the pictures mentioned above.

Based on the previous results, including what is shown in the previous sections on dwarf galaxy haloes, we select a set of parameters that are considered to be most promising. 
To account for the possibility of having a shallower (or mildly shallower) inner profile found in dwarf galaxies, we only choose relatively small cross sections, namely $\sigma_{0} / m = 0.01$ and 0.1 \units, and the set of velocity-dependent models used are ($-2, -2$), ($-1,-1$), ($0,0$), ($-2,0$), and ($-1,0$).

Figure \ref{fig:GC-profiles} overlays the promising cases of 2cDM profiles on observational data. 
Although our simulation data do not allow us to probe as deep the inner-radial range as the observational data, we are resolving enough range to see the characteristic radii where the turnover occurs for some of the 2cDM models. 
Overall, the cases presented here with $\sigma_{0}/m = 0.01$ \units follow a similar trend with the CDM except that they show mildly shallower profiles towards smaller $r/R_{\rm vir}$, which are well within the observationally inferred range.   
Those cases with $(X,0)$ with $\sigma_{0}/m = 0.1$ \units (dashed curves), on the other hand, give much larger core radii with smaller central density. This implies that even with a possible presence of baryon, which would induce a deeper gravitational potential in the cluster center, the cross-section value greater than 0.1 \units would likely to fail to conform to the observations.

As a more quantitative way to study the profiles, we show some of the cases with the mean radial density profile with 1$\sigma$ spread (shaded) and the fit (solid $\&$ dashed curves) in Figure \ref{fig:GCprofile}. Also shown in each panel is the mean profile of the CDM model (red dotted) for comparison. 
Having shown that the 2cDM model creates cored density profiles in dwarf galaxy and MW sized haloes (Section \ref{sec:dwarfs}, Paper I $\&$ II), it might be natural to consider a cored profile, such as gISO profile, for cluster density profiles as well. However, since the observed cluster data  that we want our data to be compared with can be described by either an NFW or its modified version of the generalized NFW (gNFW) \citep{newman2011, newman2013a, newman2013b, meneghetti2014, umetsu2016}, we fit our cluster sample with the gNFW model written as  
\begin{equation} \label{eq:gNFW}
\rho (r) = \frac{\rho_{s}}{ (r/r_{s})^{\tilde{\beta}} (1 + r/r_{s})^{3- \tilde{\beta} }   },
\end{equation}
where $\rho_{s}$ and $r_{s}$ are the characteristic density and radius, respectively, and $\tilde{\beta}$ is the logarithmic inner slope which adds statistical flexibility to the fitting model as opposed to that of the constant value of $\tilde{\beta} = 1$ for the CDM model. Note that the gNFW is effectively reduced to the NFW if $\tilde{\beta} \rightarrow 1$. 

We found that with the chosen set of parameters the 2cDM model can successfully create both a shallower and an NFW-like inner profile, and the gNFW model gives a reasonable fit. 
The other implications are the following:
(i) The ($X$,0) models, where $X = \{-2, -1, 0\}$, produce mild to relatively strong reduction on the inner mass density with $\sigma_{0}/m = 0.01$ and 0.1 \units. 
(ii) The symmetric cases of $(-2,-2)$ and $(-1,-1)$, which we only tested with $\sigma_{0}/m = 0.01$ \units, show a somewhat weaker effect on the density reduction in the inner-most part compared to the ($X$, 0) counterparts. Their gNFW fit also turned out nearly identical to that of the CDM model. Note that although not shown, the cases with ($-2,-1$) and ($0,-1$) are also expected to show a similar behavior with ($-1,-1$) with minor differences for GCs as it was indicated in dwarf galaxy haloes (see Figure \ref{fig:dwarf_profile_all_orig4}). 

The relatively strong effect seen in the ($X$, 0) models with $\sigma_{0}/m = 0.1$ \units implies a larger cross-section value of $\sigma_{0}/m = 1$ \units or greater for those models would likely produce a much shallower inner profile with a larger core radius ($\gtrsim 100$ kpc), thereby it could potentially conflict with observations even with the presence of baryons since the domination of baryons by mass in GCs does not extend beyond 100 kpc.
In the meantime, any values in the range of $0.01 \lesssim \sigma_{0}/m \lesssim 0.1$ \units within those models can be plausible, given that we do not consider baryonic effects.



\subsection{Fitting parameters}
\subsubsection{ $\tilde{\beta}$ vs. $r_{s}$} 

One of the primary outcomes of imposing inelastic mass conversion along with the elastic scattering to a DM model is creation of a shallower inner slope of DM halo density profile. The $\tilde{\beta}$ parameter obtained from the gNFW fitting model quantifies any deviation of the inner slope from a cuspy one, and thus provides us a quantitative measure on the strength of the effect of the 2cDM physics. By evaluating $\tilde{\beta}$ in comparison with observations, it would then provide another way of constraining the parameters used in the 2cDM model. Here we study the correlation between the two fitting parameters, $\tilde{\beta}$ and $r_{s}$, and discuss the implications.

\begin{figure}
  \centering
  \includegraphics[scale = 0.39]{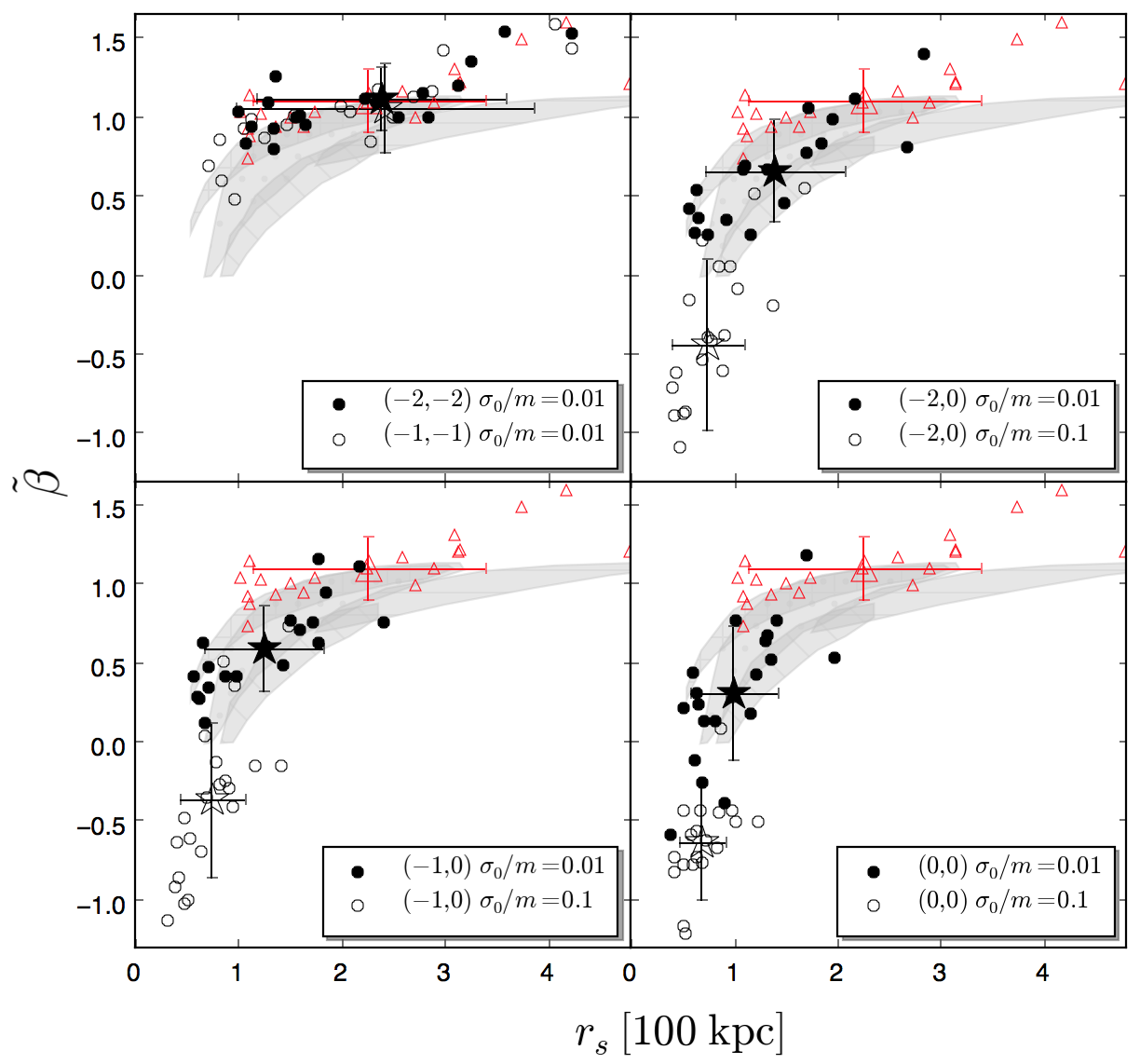}  
  \caption[]{\label{fig:beta_rs} %
Correlation between $\tilde{\beta}$ and $r_{s}$ for the selected set of parameters of 2cDM, CDM (red), and observations (shaded). Each circle (2cDM) and triangle (CDM) represents individual halo from our sample. The large stars and triangle with error bars are the mean and the 1$\sigma$ spread. 
  }
\end{figure}

Figure \ref{fig:beta_rs} shows $\tilde{\beta}$ vs. $r_{s}$ from our sample overlaid the observational data from \citet{newman2013a}. 
The selected observational data consist of A383, A611, A963 and MS2137 (1$\sigma$ confidence region) with their mean virial mass ranging from $\sim$log$ M_{200} / M_{\odot} = 14.5$ to 14.9, which is by a factor of $\sim$7 larger than the mean of our sample halo mass, but our largest halo differs from them by a factor of only less than 2. For comparison purposes, we also show the case with CDM (red triangles) in all the panels. 

We confirm that the CDM model is well within the observational range with the mean $\tilde{\beta} \sim 1$, consistent with previous studies that an NFW function can describe the observed GC density profile reasonably well. Closely following the CDM trend is the symmetric cases of $(-2,-2)$ and $(-1,-1)$ with $\sigma_{0}/m = 0.01$ \units (upper left panel). This in turn implies that with those particular set of parameters on $a_{s}$ and $a_{c}$, $\sigma_{0}/m = 0.01$ \units is hitting the lower-limit and any smaller cross section would yield results that are no more different than the collisionless CDM model.
The ($X,0$) models, on the other hand, show a clear deviation from the CDM, and the larger cross-section value yields $\tilde{\beta}$ much less than 1, corresponding to a shallower inner density profile.
A particularly strong flattening of the inner slope is clearly seen for the cases with $\sigma_{0}/m = 0.1$ \units where $\tilde{\beta}$ drops below 0, although we note that our results are still in agreement with observations within the 2$\sigma$ confidence level (not shown). 
In the meantime, the cases with $(X,0)$ $\sigma_{0}/m = 0.01$ \units show the mean value of the logarithmic inner slope $\langle \tilde{\beta} \rangle \sim 0.5$ and are well within the observed data. We caution, however, that the goodness of the gNFW fit is being compromised for these cases with $\sigma_{0}/m = 0.01$ \units for all the ($X$, 0) models due to the lack of spatial resolution in the inner-most region of the halo. That is, the density profiles shown in Figure~\ref{fig:GCprofile} imply the actual value of $\tilde{\beta}$ should be smaller than what we have obtained from the fit.
Lastly, there is a minimal difference between $(-2,0)$, $(-1,0)$ and ($0,0$), and our results indicate that the difference is simply due to statistical in nature.

A further implication on the cross section is that a larger value of $\sigma_{0} / m \gtrsim 0.1$ \units is likely to produce a large core that is inconsistent with observations, and therefore might be excluded from the plausible parameter space in the 2cDM model. This is consistent with the numerical results presented in Paper I $\&$ II that for $N$-body simulations that the 2cDM model seems to consistently reproduce agreement with observations over the many decades of halo mass most well with $\sigma_{0} / m \lesssim 0.1$ \units, regardless of the choice of $a_{s}$ and $a_{c}$.
We argue that this constrain would remain plausible even with the possibility of including baryonic physics in our simulations.

\subsubsection{Concentration parameter} \label{sec:concentration}

For consistency check, we also examined the concentration parameter of our cluster sample with observations. The concentration parameter is defined in terms of the virial quantity as $c_{200} \equiv R_{200} / r_{s}$, which describes the halo concentration derived from an NFW profile. The parameter naturally appears in both an NFW and gNFW profiles, and hence they can be measured and compared with observations. Both observational and theoretical studies have shown that the concentration can be dependent on the halo mass and the redshift or its assembly history, both in a form of declining power-laws over a given halo mass range \citep[e.g.][]{bullock2001, buote2007, schmidt2007, maccio2008, duffy2008, okabe2010, oguri2012}. 

Such power-law dependency could however give us an overestimate of the DM annihilation flux signal (or $\gamma$-ray detection signal) expected from the highly concentrated substructures, and more modest substructure boosts are expected from a much smaller mass scale \citep{sanchez-conde2014}.
It has been raised that there may be a tension between the observed concentration and the one from numerical simulations, in which the former appears to have some factors larger concentration than that of simulations \citep{broadhurst2008, oguri2009}, whereas some studies found otherwise \citep{merten2015, sereno2015, umetsu2016}. Here we briefly summarize the results on the concentration, or more specifically the concentration-mass (c-M) relation, for the 2cDM and CDM.

From the gNFW fit, we found that the case with $(-2,0)$ $\sigma_{0}/m = 0.1$ \units yields the mean concentration from the sample is $\langle c_{200} \rangle = 15.5 \pm 6.0$, whereas that of CDM is $\langle c_{200} \rangle = 5.1 \pm 2.1$ with 1$\sigma$ error. The cases with $(-2,-2)$ and $(-1,-1)$ $\sigma_{0}/m = 0.01$ \units are only marginally different from CDM. 
In general, the value of the mean concentration for the case with a larger cross section turned out much smaller when an NFW is used for the fit. For example, $(-2,0)$ $\sigma_{0}/m = 0.1$ \units gives $\langle c_{200} \rangle \sim 2 $, which is roughly a factor of 7 smaller than the value from gNFW. The large discrepancy is attributed to the poorer fit given by NFW compared to the gNFW for profiles that have a shallower inner density slope, and thus a similar but more mild discrepancy is seen in the cases with $(-1,0)$ and $(0,0)$ $\sigma_{0}/m = 0.1$ \units. 
In the meantime, for the cases with smaller cross section of $\sigma_{0}/m = 0.01$ \units are within 1$\sigma$ from each other between gNFW and NFW.

We found that there is an inconsistent trend in the values of $r_{s}$, hence $c_{200}$, when compared between the gNFW and NFW in the 2cDM results, especially for the cases with a large cross-section value of 0.1 \units for $(X,0)$. While for the NFW fit, a larger cross section yields a larger scale radius $r_{s}$ with a smaller concentration $c_{200}$, the gNFW gives the opposite trend with a larger cross section producing a smaller $r_{s}$ and a higher $c_{200}$.  
Even though the goodness of the fit in terms of the reduced $\chi^{2}$ value does not differ significantly between the two profile models (especially true if the profile shape is close to that of CDM), the gNFW profile captures the mildly shallower or flat inner part of density profile better. 
In other words, the gNFW is more sensitive in determining the turnover of the profile, which is where $r_{s}$ is essentially defined. 
For the 2cDM model, the DM mass are re-distributed and pushed outward after the mass conversion interactions take place, resulting in the shift of the position of $r_{s}$ in the density profile and creating a more sudden turnover compared to a more smooth transition seen in an NFW profile.
The inconsistent trend found in the concentration from NFW and gNFW can thus mostly be due to (i) the inability of the NFW profile model to accurately determine $r_{s}$ for a flat profile and (ii) the gNFW can be too sensitive to the more drastic turnover of a flat 2cDM density profiles.

Due to the limited statistical sample over the range of halo mass and the GC counts, we do not attempt to fit our data on the c-M relation with a power-law. The relatively large scatter among the sample also prevents us from drawing any firm conclusion on the anti-correlation of the c-M relation seen in literature. 
Our results however highlights that in terms of the concentration parameter, a cross-section value of $\sigma_{0}/m = 0.01$ \units in the 2cDM model, especially for the cases with $(-2,-2)$ and $(-1,-1)$, yields good agreement with the CDM. The only minute difference from the CDM is that there is an implication from the density profile that the inner-most slope ($r \lesssim 10$ kpc) is shallower. 
If there is a better numerical resolution to resolve the inner radial region, then the fitting parameter obtained from the gNFW would have been slightly affected and possibly producing a slightly larger concentration than the CDM. It is inconclusive whether the tension between the numerical/theoretical predictions and observations can be explained by the 2cDM.





\section{Discussion} \label{sec:discussion}

The results shown in this work do not consider the baryonic physics. For DM-dominated systems this is a reasonable assumption in general, at least to test and constrain a DM model for our purposes. 
However, both observations and numerical simulations have shown that even for DM-dominated systems the baryonic physics plays a role in certain regimes, although the significance of the effect may depend on the assumptions and models at hand. 
In this section we discuss the implications from this work and the possible effects of including baryonic physics combined with 2cDM physics on dwarf and GC systems.

\subsection{Implications on baryonic effects}

\subsubsection{Dwarf galaxies} 
  
Dwarf galaxies are known to host relatively small fraction of stars and gas (high mass-to-light ratios) and mostly dominated by DM mass. The inclusion of baryons in our analysis would therefore less likely affect significantly the overall shape of the 2cDM density profiles shown in this work. However, unlike galaxy clusters, they are formed in the early Universe via the bottom-up structure formation scenario.
This requires us to examine how the 2cDM physics plays a role in terms of the halo evolutionary processes.  
To quantitatively check this, we examined the evolution of the fitting parameters from the gISO profile, namely $r_{c}$, $\rho_{c}$ and $p$ (in Eq. (\ref{eq:gISO})), over the scale factor of $0.25 \leq a \leq 1$ ($0 \leq z \leq 3$). We found that the evolution of $\rho_{c}$ follows a power-law with the logarithmic slope of $d \log \rho_{c} / d\log a \sim 1.4$, which is nearly independent of whether the elastic scattering, inelastic mass conversion, or both are assumed in the 2cDM model with $(-2,-2)$. The slope is also insensitive to the cross section, at least for smaller ones ($\sigma_{0}/ m = 0.001$ and 0.01 \units). The evolution of $p$ as a function of scale factor also shows a modest power-law relation: it indicates the haloes can be better described as isothermal at higher redshift $z = 3$ and the halo deviates from it gradually towards the current time. Importantly, we also found that as oppose to the gradual increase of $\rho_{c}$, the core radius steadily decreases towards the current time. This implies the central region of dwarf haloes can be  less concentrated at earlier redshift (as early as $z = 3$), especially so if the mass conversion is enabled, and that in this scenario it is likely that the formation of gas and stars in such lower dense environment at earlier time could delay the burst of star formation significantly. 

The FIRE hydrodynamical simulations \citep{Onorbe2015} showed that a bursty stellar feedback can create a DM density core size of $\sim 1$ kpc in the inner-most region of dwarf galaxies only above a stellar mass of M$_{star} \sim 10^{6.3}$ M$_{\odot}$, depending on the star formation histories.
Similarly, \citet{governato2012} found the inefficiency in the transfer of stellar feedback energy to DM in the system below the virial halo mass of $M_{vir} < 5 \times 10^{9}$ M$_{\odot}$ to soften the cuspy DM density profile. In this work we showed that the 2cDM model can create a sizable DM density core even in the virial halo mass of as small as $M_{vir} \sim 10^{7 - 8}$ M$_{\odot}$, which is up to a few orders of magnitude smaller than their counterparts, and that our results show the mass conversion naturally creates a core without relying on baryonic feedback. 

\subsubsection{Galaxy clusters}

Possible baryonic effect in GCs can particularly be noticeable in the core of clusters where the complex interplay among the central galaxies, hot bubbles, cold stream, etc. is not well understood \citep[e.g.][]{mcnamara2007}. However it has been observed that the central region (as small as $r \sim 10$ kpc) is dominated by stellar mass, and hence the total density profile (luminous + DM) has a logarithmic inner slope steeper than that of an NFW \citep[e.g.][]{sand2004, newman2013a, newman2013b}. CDM-based numerical simulations generally confirm this picture.  
Meanwhile, the so-called overcooling problem has also been well-known in the numerical simulations to cause condensation of baryonic matter in the deep gravitational potential, which induces an exceeding amount of cold gas in the cluster center, resulting in excessive star formation \citep[e.g.][]{borgani2011}. 
This is generally attributed to the inefficiency of the baryonic feedback processes, namely Active Galactic Nuclei (AGN) feedback, of which its strength and efficiency can be controlled by the assumptions made in the model and the parameters associated with them, to counteract with the cluster's gravity within the CDM framework.
In fact, numerical simulations with an AGN feedback have predicted both a cuspy \citep{schaller2015} or shallower DM inner density profile \citep{martizzi2012, martizzi2013}, in which the latter can primarily be created by a strong AGN feedback. 
That is, the feedback can be energetic enough to quench late star formation and turn the cuspy DM density profile to a flat one by means of causing a strong perturbations in the gravitational potential and removing the DM mass from the central part of clusters.

Due to the nature of the 2cDM model to create a shallower gravitational potential in the halo center, producing a flat core in the DM density profile, the model naturally alleviates the so-called overcooling problem seen in the CDM without relying on the baryonic feedback.
In fact, for the cases with $\sigma_{0} / m = 0.01$ \units tested in this work, creation of a core size of roughly $\lesssim$ 30 kpc is evident, while with $\sigma_{0} / m = 0.1$ \units they are $\sim 60$ kpc (note that these values do not necessarily correspond to the characteristic radius $r_{s}$ measured from an NFW or gNFW profile). In other words, the DM density within such radial range is noticeably reduced and the gravitational potential can significantly be shallower compared to that of a CDM halo. 
The immediate impact is a suppression of overly concentrated cold gas in the core, thus it follows that it could inhibit the excessive star formation. Such effect is expected to be particularly significant for cases with $\sigma_{0}/m \gtrsim 0.1$ \units within the 2cDM paradigm.
In the meantime, combining a strong AGN feedback as described in \citet{martizzi2013} with the 2cDM model would likely create a core size that is larger than what is observed, especially for the cases with $\sigma_{0}/m \gtrsim 0.1$ \units or greater. This would certainly worsen the discrepancy with observations.

The effect of including baryons on the c-M relation has also been studied in literature and implied to have a non-negligible impact \citep[e.g.][]{fedeli2012}. This however is not trivial because the measurement of the concentration requires accurate determination of the characteristic radius $r_{s}$ that is dependent on how well a gNFW or NFW model fits the profile. In Section \ref{sec:concentration} we showed that the value of $r_{s}$ starts to deviate from each other in between gNFW and NFW for the case with larger cross-section values, mostly owning to the lack of accuracy in NFW to capture the shallower inner slope. 
We argue that although our results are inconclusive on whether the apparent discrepancy found by some studies in the c-M relation can be explained by the 2cDM model, inclusion of baryonic physics would unlikely transform the inner DM density profile to be an even shallower one, unless a strong AGN feedback is employed. However, additional presence of baryon concentration induced by the gas cooling and the presence of large stellar mass in the central region could enhance the DM concentration in that region through gravitational attractions, which would help make some of the flat 2cDM profiles more towards that of a cuspy CDM-like profile. Therefore, there lies no problem since some observational studies found that the observed c-M relation agrees with that of CDM predictions.

\subsection{Constrains from cluster mergers}

Cluster merger has been studied widely and of great importance in establishing a firm evidence of DM existence \citep{clowe2004}. It has also been used to constrain the self-interacting nature of DM based on the offset between the collisionless stellar component and the DM component, measured from optical images and gravitational lensing data. 
The well-known Bullet Cluster shows the gas distribution detected in the optical or X-ray images lags behind the collisionless stars and DM \citep{markevitch2004}, which signifies that DM cannot be fluid-like or any more than modestly collisional. The other merging clusters were also studied to set a constrain on the self-interacting nature of the DM. Similar to the Bullet Cluster, their inferred self-interaction cross section per unit mass ($\sigma /m $) based on the scattering depth of the DM, $\tau_{DM} = (\sigma/m) \Sigma_{DM}$, where $\Sigma_{DM}$ is the DM surface mass density estimated from lensing data, has been reported to be $\sigma/m \lesssim \mathcal{O}(1)$ \units as an order of magnitude estimate for the upper limit \citep{markevitch2004, bradac2008, merten2011, dawson2012, clowe2012, harvey2015}. Theoretical and numerical simulations of cluster merger have also reached a similar constrain \citep[e.g.][]{randall2008, kahlhoefer2014,robertson2017}. 
The inferred cross-section value of $\sigma_{0} / m \lesssim 0.1 $ \units as the 2cDM model's preferred value is clearly in reasonable agreement with those cluster merger studies. 

\begin{table*} 
\centering
\tabcolsep=0.08cm
\begin{tabular}{ccccc|c|ccc|ccc}
\hline\hline
 & & MW &  &  & Dwarf & GC &  &  & Theoretical  \\
Model & $\sigma_{0}/m$ & Density profile & VF &  RHDF & Density Profile & Density Profile & $\tilde{\beta}$-$r_{s}$ & $c$-$M$ relation & preference  \\
\hline

$(-2,-2)$  &  0.001    & Maybe     & 	YES  & YES  &    NO  & --  & --  & -- & YES\\
		& 0.01 & Maybe &  Maybe & YES & YES  & YES & YES & YES  & YES\\
               &  0.1 & YES     &	YES   & YES &  NO  &  --  & -- &  -- & YES \\
 	      &  1    & YES     & 	YES  & YES  &    NO  & --  & --  & -- & YES\\
	      &  10    & NO     & 	YES  & YES  &    NO  & --  & --  & -- & YES\\
\hdashline
$(-1,-2)$  &  0.001    & Maybe     & 	YES  & YES  &    NO  & --  & --  & -- & \\
	   & 0.01 & Maybe & YES & YES & NO & -- & -- & -- & \\
           &  0.1    &  YES &	YES     & YES &  NO  & -- & -- & -- &  \\
 	   &  1     &  YES   & 	YES     &  YES &   NO   & --  &  -- & -- & \\
	   &  10    & NO     & 	YES  & YES  &    NO  & --  & --  & -- & \\
\hdashline
$(0,-2)$  &  0.001    & Maybe     & 	YES  & YES  &    NO  & --  & --  & -- & \\
	   & 0.01 & Maybe  & YES & YES & NO & -- & -- & -- & \\
           &  0.1 &  YES & YES	   & YES &  NO  &  --  & -- & -- &  \\
 	   &  1  &  YES   & YES	  &  YES &   NO   &  -- & --  & -- & \\
	   &  10    & NO     & 	YES  & YES  &    NO  & --  & --  & -- & \\
\hdashline
$(-2,-1)$ &  0.001    & Maybe     & 	YES  & YES  &    NO  & --  & --  & -- & \\
	   & 0.01 & Maybe  & Maybe & YES & YES & -- & -- & -- & \\
           &  0.1 & YES  & YES	   & YES      &   NO & --  & -- & -- &  \\
 	   &  1  &  YES  &  YES &       YES    & NO    &  -- & --  & -- & \\
	   &  10    & NO     & 	YES  & YES  &    NO  & --  & --  & -- & \\
\hdashline
$(-1,-1)$ &  0.001    & Maybe     & 	YES  & YES  &    NO  & --  & --  & -- & YES\\ 
	  & 0.01 & Maybe & Maybe & YES & YES & YES & YES & YES & YES \\
           &  0.1 & YES & YES              & YES &  NO &   -- & -- & -- & YES \\
 	   &  1  & YES  & YES              & YES &   NO & -- & -- & -- & YES \\
	   &  10    & NO     & 	YES  & YES  &    NO  & --  & --  & -- & YES\\
\hdashline
$(0,-1)$ &  0.001    & Maybe     & 	YES  & YES  &    NO  & --  & --  & -- & \\ 
	  & 0.01 & Maybe  & Maybe & YES & YES & -- & -- & -- & \\
           &  0.1 & YES & YES  &     YES      &  NO  &  -- & -- & -- &  \\
 	   &  1  & YES  & YES  &    YES  &   NO     &  -- & --  &--  & \\
	   &  10    & NO     & 	YES  & YES  &    NO  & --  & --  & -- & \\
\hdashline
$(-2,0)$ &  0.001    & Maybe     & 	NO  & YES  &    NO  & --  & --  & -- & \\ 
	  & 0.01 & Maybe & NO  &  NO & YES & YES & YES & YES & \\
           &  0.1 & YES &	Maybe   & YES &  YES  & Maybe & YES & YES  & \\
 	   &  1  &  YES &  YES       &  NO    &    NO  & --  & --  & -- & \\
	   &  10    & NO     & 	NO  & YES  &    NO  & --  & --  & -- & \\
\hdashline
$(-1,0)$ &  0.001    & Maybe     & 	NO  & YES  &    NO  & --  & --  & -- & \\ 
	   & 0.01 & Maybe   & NO & NO  & YES & YES & YES & YES &  \\
           &  0.1 &  YES &	Maybe   & YES &   YES &  Maybe & YES & YES & \\
 	   &  1  &  YES  & YES	  &  NO &    NO  &  -- & --  & -- & \\
	   &  10    & YES     &  NO  & YES  &    NO  & --  & --  & -- & \\
\hdashline
$(0,0)$ &  0.001    & Maybe     & NO  & YES  &    NO  & --  & --  & -- & YES\\ 
	  & 0.01 & YES & NO &       NO   & YES & YES & YES & YES & YES\\
           &  0.1 & YES &	Maybe   & YES &  YES  &  Maybe  & YES & YES & YES \\
 	   &  1  &  YES & YES  &   NO   &   NO   & --  &  -- & -- & YES\\
	   &  10    & NO    & 	NO  & YES  &    NO  & --  & --  & -- & YES\\
\hdashline
SIDM  &  0.001    & YES     & 	NO  & --  &   --  & --  & --  & -- & \\
	   & 0.01 & YES & NO &       --   & -- & -- & -- & -- & \\
           &  0.1 & YES &	NO   & -- &  --  &  --  & -- & -- &  \\
 	   &  1  &  YES & NO  &   --   &  --   & --  &  -- & -- & \\
	   &  10    &   --   & -- & --  &   --  & --  & --  & -- & \\
\hdashline
CDM & -- & Maybe & Maybe  & Maybe & Maybe & YES & YES & YES & \\[1 ex]              
\hline\hline

\end{tabular}
\caption[]{ Yes-and-No table summarizing the general compatibility of each model to observations. Results on Milky Way (MW) haloes are from Paper I $\&$ II, and the rest are from this work. YES implies the set of parameter reproduces a consistent result with observations, while NO indicates otherwise. 'Maybe' is for the inconclusive cases where additional work such as with baryonic physics would be required. All SIDM cases are $(-2,-2)$-based with the inelastic mass conversion disabled. The cases with $(-4, X)$ where $X = -2, -1, 0$ are omitted here since they were shown to be disfavored in the previous work. }
\label{table:yesno}
\end{table*}



\section{Summary and Conclusions} \label{sec:CN}

In this work we explored the effect of the 2cDM physics on the DM haloes of the size of hosting dwarf galaxies and clusters of galaxies, effectively covering 7 orders of magnitudes in the virial halo mass. Following the studies on the MW-sized haloes presented in Paper I $\&$ II, the results presented in this work places a more stringent constrain on the 2cDM model parameters based on the $N$-body cosmological simulations. 

Constraining parameters is one of the major goals of this work, and we first tested the promising set of parameters on dwarf haloes. Those promising parameters that were inferred from the results of MW-sized halo studies presented in Paper I $\&$ II include the symmetric cases of $(a_{s}, a_{c}) = (-2,-2), (-1,-1)$, and $(0,0)$, and others with $a_{s}, a_{c} = -2, -1,$ and 0, totaling 9 sets of DM cross section's velocity dependent or independent models. Note that models with $\sigma(v) \propto 1/v^{4}$ (i.e., either $a_{s}$ or $a_{c} = -4$) were not considered in this work due to the fact that such strong velocity dependence puts DM close to reaching the fluid regime with characteristically high interaction rates within the reasonable choice of cross-section values. The other key parameter chosen as a fiducial value for both dwarf and GC is the kick velocity $V_{k} = c\sqrt{2 \Delta m / m} = 100$ km s$^{-1}$ which accounts for the mass degeneracy between the two mass eigenstates as $\Delta m / m \sim 10^{-8} - 10^{-7}$. 
Additionally, we explored the 4 decades of DM cross-section values ranging from $\sigma_{0}/m = 0.001$ to 1 \units, where we excluded a case with 10 \units for the similar reason to the case with $\sigma(v) \propto 1/v^{4}$. By examining the halo structures of the top 5 most well resolved haloes in our sample with their virial mass of $\sim 10^7 - 10^8$ M$_{\odot}$, we found the following:
\begin{itemize}
\item $\sigma_{0}/m \gtrsim 0.1$ \units are generally disfavored for all the models except for $(a_{s}, a_{c}) = (X, 0)$ where $X = -2, -1, 0$. The cross section can be as small as $\sigma_{0}/m = 0.001$ \units for $(-2,-2)$, $(-2,-1)$, $(-1,-1),$ and ($0, -1$) to show at least a modest deviation from the cuspy NFW profile in the logarithmic inner slope. However, with such a small cross section $(X,0)$ makes little difference and its halo profile is nearly identical to the CDM counterpart. From this, our results indicate that the minimum cross-section value to make some noticeable flattening of the inner density profile lies somewhere between $0.001 \lesssim \sigma_{0}/m \lesssim 0.01$ \units in the 2cDM model. 
\item The mass loss fraction due mainly to the inelastic mass conversion is particularly profound within $r < 300$ pc compared to that of within the virial radius $r < R_{vir}$ (Figure \ref{fig:M2cDM_by_Mcdm}). On average a 2cDM halo with (0,0) can lose $\sim 10 \%$ for $\sigma_{0}/m = 0.01$, $\sim 30 \%$ for $\sigma_{0}/m = 0.1$, and $\sim 90 \%$ for $\sigma_{0}/m = 1$ \units of DM mass from $r < 300$ pc in comparison with the CDM counterpart. 
\end{itemize}

Following from what is implied in the dwarf haloes, we chose only a limited set of parameters which are considered to be some of the most promising to further test the 2cDM model on the GCs. The selected parameters are the symmetric cases of $(-2,-2), (-1,-1), (0,0)$ and a few asymmetric cases of $(-2,0)$ and $(-1,0)$. The cross section was chosen to be either 0.1 or 0.01 \units, excluding 1 \units which is unlikely to be plausible. We studied a cluster sample of $\sim 20$ taken at $z = 0.25$ over the halo mass ($M_{200}$) range of $10^{14} - 10^{15}$ M$_{\odot}$ and performed a fit on the density profile with the gNFW and NFW radial profiles and examine the fitting parameters by comparing with observational data. The key findings are as follows:
\begin{itemize}
\item $\sigma_{0}/m = 0.01$ \units is well within the observations, which can create density profiles that are not too dissimilar to that of CDM but with a slightly shallower, less cuspy inner slope. Thus, with such cross section the 2cDM could explain both possibilities of creating an NFW-like and a shallower profile with a reasonable core size of $\lesssim 30$ kpc. Presence of baryonic mass in the core region would also be compatible to observations, for the dominance of stellar mass in the central galaxies typically does not exceed a few tens of kpc measured from the halo center. However, we note that a possible effect of strong AGN feedback in the central region could compromise the agreement with observations by creating unrealistically large core size. 
\item The concentration parameters derived from the gNFW and NFW profiles show generally reasonable agreement with both observations and CDM-based numerical predictions. We found some degree of deviation from CDM for cases with a larger cross section of $\sigma_{0}/m = 0.1$ \units, but the resulting concentration can still be within the error due to the relatively large scatter within the sample.
\end{itemize}

For both dwarf and GC simulations performed in this work, we have enough spatial resolution to probe the radial scale of our interest to discern whether a certain set of parameters should be ruled out. While some parameters are shown to be inconsistent with observations and can be ruled out, there remains a handful of them that can still be a possibility, even when the baryonic physics is considered. We summarize the entire list of 2cDM parameters that are either tested or implied in this work and from Paper I $\&$ II in Table \ref{table:yesno}. By considering MW, dwarf, GC and possible effects from including baryonic physics, the 2cDM's most preferred cross-section value is $\sigma_{0}/m \lesssim 0.1$ \units, which is in agreement with both observations and theoretical/numerical predictions on self-interacting DM models \citep[e.g.][]{peter2013, rocha2013}. We also found that the symmetric models of $(a_{s}, a_{c}) = (-2,-2), (-1,-1)$ and $(0,0)$ generally work well to reproduce desirable results. The implication is that with the inelastic mass conversion, the model allows both a strong velocity-dependent and velocity-independent cross section as a possibility. To further investigate the model, a set of dark matter + baryonic hydrodynamics simulations are required with a better statistical sample than what is presented in this work. 

%


\section*{Acknowledgements}
The simulations for this work were carried out at the Advanced Computing Facilities at the University of Kansas. MM is grateful to the the Institute for Theory and Computation at Harvard University and the Kavli Institute for Theoretical Physics for support and hospitality. This research was supported in part by the NSF under grant No. PHY-1748958, the DOE under grant No. DE-SC0016368 and the DOE EPSCOR under grant No. DE-SC0019474.
\bibliographystyle{mnras} 
\bibliography{2cDM-paper3}    

\begin{thebibliography}{}
\makeatletter
\relax
\def\mn@urlcharsother{\let\do\@makeother \do\$\do\&\do\#\do\^\do\_\do\%\do\~}
\def\mn@doi{\begingroup\mn@urlcharsother \@ifnextchar [ {\mn@doi@}
  {\mn@doi@[]}}
\def\mn@doi@[#1]#2{\def\@tempa{#1}\ifx\@tempa\@empty \href
  {http://dx.doi.org/#2} {doi:#2}\else \href {http://dx.doi.org/#2} {#1}\fi
  \endgroup}
\def\mn@eprint#1#2{\mn@eprint@#1:#2::\@nil}
\def\mn@eprint@arXiv#1{\href {http://arxiv.org/abs/#1} {{\tt arXiv:#1}}}
\def\mn@eprint@dblp#1{\href {http://dblp.uni-trier.de/rec/bibtex/#1.xml}
  {dblp:#1}}
\def\mn@eprint@#1:#2:#3:#4\@nil{\def\@tempa {#1}\def\@tempb {#2}\def\@tempc
  {#3}\ifx \@tempc \@empty \let \@tempc \@tempb \let \@tempb \@tempa \fi \ifx
  \@tempb \@empty \def\@tempb {arXiv}\fi \@ifundefined
  {mn@eprint@\@tempb}{\@tempb:\@tempc}{\expandafter \expandafter \csname
  mn@eprint@\@tempb\endcsname \expandafter{\@tempc}}}

\bibitem[\protect\citeauthoryear{{Allen}, {Evrard}  \& {Mantz}}{{Allen}
  et~al.}{2011}]{allen2011}
{Allen} S.~W.,  {Evrard} A.~E.,   {Mantz} A.~B.,  2011, \mn@doi [\araa]
  {10.1146/annurev-astro-081710-102514}, \href
  {http://adsabs.harvard.edu/abs/2011ARA%26A..49..409A} {49, 409}

\bibitem[\protect\citeauthoryear{{Arraki}, {Klypin}, {More}  \&
  {Trujillo-Gomez}}{{Arraki} et~al.}{2014}]{arraki2014}
{Arraki} K.~S.,  {Klypin} A.,  {More} S.,   {Trujillo-Gomez} S.,  2014, \mn@doi
  [\mnras] {10.1093/mnras/stt2279}, \href
  {http://adsabs.harvard.edu/abs/2014MNRAS.438.1466A} {438, 1466}

\bibitem[\protect\citeauthoryear{{Bartelmann} \& {Schneider}}{{Bartelmann} \&
  {Schneider}}{2001}]{bartelmann2001}
{Bartelmann} M.,  {Schneider} P.,  2001, \mn@doi [\physrep]
  {10.1016/S0370-1573(00)00082-X}, \href
  {http://adsabs.harvard.edu/abs/2001PhR...340..291B} {340, 291}

\bibitem[\protect\citeauthoryear{{Binney} \& {Tremaine}}{{Binney} \&
  {Tremaine}}{2008}]{binney2008}
{Binney} J.,  {Tremaine} S.,  2008, {Galactic Dynamics: Second Edition}.
Princeton University Press

\bibitem[\protect\citeauthoryear{{Borgani} \& {Kravtsov}}{{Borgani} \&
  {Kravtsov}}{2011}]{borgani2011}
{Borgani} S.,  {Kravtsov} A.,  2011, \mn@doi [Advanced Science Letters]
  {10.1166/asl.2011.1209}, \href
  {http://adsabs.harvard.edu/abs/2011ASL.....4..204B} {4, 204}

\bibitem[\protect\citeauthoryear{{Boylan-Kolchin}, {Bullock}  \&
  {Kaplinghat}}{{Boylan-Kolchin} et~al.}{2011}]{boylan-kolchin2011}
{Boylan-Kolchin} M.,  {Bullock} J.~S.,   {Kaplinghat} M.,  2011, \mn@doi
  [MNRAS] {10.1111/j.1745-3933.2011.01074.x}, \href
  {http://adsabs.harvard.edu/abs/2011MNRAS.415L..40B} {415, L40}

\bibitem[\protect\citeauthoryear{{Brada{\v c}}, {Allen}, {Treu}, {Ebeling},
  {Massey}, {Morris}, {von der Linden}  \& {Applegate}}{{Brada{\v c}}
  et~al.}{2008}]{bradac2008}
{Brada{\v c}} M.,  {Allen} S.~W.,  {Treu} T.,  {Ebeling} H.,  {Massey} R.,
  {Morris} R.~G.,  {von der Linden} A.,   {Applegate} D.,  2008, \mn@doi [\apj]
  {10.1086/591246}, \href {http://adsabs.harvard.edu/abs/2008ApJ...687..959B}
  {687, 959}

\bibitem[\protect\citeauthoryear{{Broadhurst}, {Umetsu}, {Medezinski}, {Oguri}
  \& {Rephaeli}}{{Broadhurst} et~al.}{2008}]{broadhurst2008}
{Broadhurst} T.,  {Umetsu} K.,  {Medezinski} E.,  {Oguri} M.,   {Rephaeli} Y.,
  2008, \mn@doi [\apjl] {10.1086/592400}, \href
  {http://adsabs.harvard.edu/abs/2008ApJ...685L...9B} {685, L9}

\bibitem[\protect\citeauthoryear{{Brook} \& {Di Cintio}}{{Brook} \& {Di
  Cintio}}{2015}]{brook2015}
{Brook} C.~B.,  {Di Cintio} A.,  2015, \mn@doi [\mnras] {10.1093/mnras/stv864},
  \href {http://adsabs.harvard.edu/abs/2015MNRAS.450.3920B} {450, 3920}

\bibitem[\protect\citeauthoryear{{Brooks} \& {Zolotov}}{{Brooks} \&
  {Zolotov}}{2014}]{brooks2014}
{Brooks} A.~M.,  {Zolotov} A.,  2014, \mn@doi [\apj]
  {10.1088/0004-637X/786/2/87}, \href
  {http://adsabs.harvard.edu/abs/2014ApJ...786...87B} {786, 87}

\bibitem[\protect\citeauthoryear{{Brooks}, {Kuhlen}, {Zolotov}  \&
  {Hooper}}{{Brooks} et~al.}{2013}]{brooks2013}
{Brooks} A.~M.,  {Kuhlen} M.,  {Zolotov} A.,   {Hooper} D.,  2013, \mn@doi
  [\apj] {10.1088/0004-637X/765/1/22}, \href
  {http://adsabs.harvard.edu/abs/2013ApJ...765...22B} {765, 22}

\bibitem[\protect\citeauthoryear{{Bullock}, {Kolatt}, {Sigad}, {Somerville},
  {Kravtsov}, {Klypin}, {Primack}  \& {Dekel}}{{Bullock}
  et~al.}{2001}]{bullock2001}
{Bullock} J.~S.,  {Kolatt} T.~S.,  {Sigad} Y.,  {Somerville} R.~S.,  {Kravtsov}
  A.~V.,  {Klypin} A.~A.,  {Primack} J.~R.,   {Dekel} A.,  2001, \mn@doi
  [\mnras] {10.1046/j.1365-8711.2001.04068.x}, \href
  {http://adsabs.harvard.edu/abs/2001MNRAS.321..559B} {321, 559}

\bibitem[\protect\citeauthoryear{{Buote}, {Gastaldello}, {Humphrey},
  {Zappacosta}, {Bullock}, {Brighenti}  \& {Mathews}}{{Buote}
  et~al.}{2007}]{buote2007}
{Buote} D.~A.,  {Gastaldello} F.,  {Humphrey} P.~J.,  {Zappacosta} L.,
  {Bullock} J.~S.,  {Brighenti} F.,   {Mathews} W.~G.,  2007, \mn@doi [\apj]
  {10.1086/518684}, \href {http://adsabs.harvard.edu/abs/2007ApJ...664..123B}
  {664, 123}

\bibitem[\protect\citeauthoryear{{Burkert}}{{Burkert}}{2015}]{burkert2015}
{Burkert} A.,  2015, \mn@doi [\apj] {10.1088/0004-637X/808/2/158}, \href
  {http://adsabs.harvard.edu/abs/2015ApJ...808..158B} {808, 158}

\bibitem[\protect\citeauthoryear{{Clowe}, {Gonzalez}  \& {Markevitch}}{{Clowe}
  et~al.}{2004}]{clowe2004}
{Clowe} D.,  {Gonzalez} A.,   {Markevitch} M.,  2004, \mn@doi [\apj]
  {10.1086/381970}, \href {http://adsabs.harvard.edu/abs/2004ApJ...604..596C}
  {604, 596}

\bibitem[\protect\citeauthoryear{{Clowe}, {Markevitch}, {Brada{\v c}},
  {Gonzalez}, {Chung}, {Massey}  \& {Zaritsky}}{{Clowe}
  et~al.}{2012}]{clowe2012}
{Clowe} D.,  {Markevitch} M.,  {Brada{\v c}} M.,  {Gonzalez} A.~H.,  {Chung}
  S.~M.,  {Massey} R.,   {Zaritsky} D.,  2012, \mn@doi [\apj]
  {10.1088/0004-637X/758/2/128}, \href
  {http://adsabs.harvard.edu/abs/2012ApJ...758..128C} {758, 128}

\bibitem[\protect\citeauthoryear{{Col{\'{\i}}n}, {Avila-Reese}, {Valenzuela}
  \& {Firmani}}{{Col{\'{\i}}n} et~al.}{2002}]{colin2002}
{Col{\'{\i}}n} P.,  {Avila-Reese} V.,  {Valenzuela} O.,   {Firmani} C.,  2002,
  \mn@doi [\apj] {10.1086/344259}, \href
  {http://adsabs.harvard.edu/abs/2002ApJ...581..777C} {581, 777}

\bibitem[\protect\citeauthoryear{{Dav{\'e}}, {Spergel}, {Steinhardt}  \&
  {Wandelt}}{{Dav{\'e}} et~al.}{2001}]{dave2001}
{Dav{\'e}} R.,  {Spergel} D.~N.,  {Steinhardt} P.~J.,   {Wandelt} B.~D.,  2001,
  \mn@doi [\apj] {10.1086/318417}, \href
  {http://adsabs.harvard.edu/abs/2001ApJ...547..574D} {547, 574}

\bibitem[\protect\citeauthoryear{{Dawson} et~al.,}{{Dawson}
  et~al.}{2012}]{dawson2012}
{Dawson} W.~A.,  et~al., 2012, \mn@doi [\apjl] {10.1088/2041-8205/747/2/L42},
  \href {http://adsabs.harvard.edu/abs/2012ApJ...747L..42D} {747, L42}

\bibitem[\protect\citeauthoryear{{Diemand}, {Kuhlen}, {Madau}, {Zemp}, {Moore},
  {Potter}  \& {Stadel}}{{Diemand} et~al.}{2008}]{diemand2008}
{Diemand} J.,  {Kuhlen} M.,  {Madau} P.,  {Zemp} M.,  {Moore} B.,  {Potter} D.,
    {Stadel} J.,  2008, \mn@doi [\nat] {10.1038/nature07153}, \href
  {http://adsabs.harvard.edu/abs/2008Natur.454..735D} {454, 735}

\bibitem[\protect\citeauthoryear{{Dubinski} \& {Carlberg}}{{Dubinski} \&
  {Carlberg}}{1991}]{dubinski1991}
{Dubinski} J.,  {Carlberg} R.~G.,  1991, \mn@doi [\apj] {10.1086/170451}, \href
  {http://adsabs.harvard.edu/abs/1991ApJ...378..496D} {378, 496}

\bibitem[\protect\citeauthoryear{{Duffy}, {Schaye}, {Kay}  \& {Dalla
  Vecchia}}{{Duffy} et~al.}{2008}]{duffy2008}
{Duffy} A.~R.,  {Schaye} J.,  {Kay} S.~T.,   {Dalla Vecchia} C.,  2008, \mn@doi
  [\mnras] {10.1111/j.1745-3933.2008.00537.x}, \href
  {http://adsabs.harvard.edu/abs/2008MNRAS.390L..64D} {390, L64}

\bibitem[\protect\citeauthoryear{{Dutton} \& {Macci{\`o}}}{{Dutton} \&
  {Macci{\`o}}}{2014}]{dutton2014}
{Dutton} A.~A.,  {Macci{\`o}} A.~V.,  2014, \mn@doi [\mnras]
  {10.1093/mnras/stu742}, \href
  {http://adsabs.harvard.edu/abs/2014MNRAS.441.3359D} {441, 3359}

\bibitem[\protect\citeauthoryear{{Elbert}, {Bullock}, {Garrison-Kimmel},
  {Rocha}, {O{\~n}orbe}  \& {Peter}}{{Elbert} et~al.}{2015}]{elbert2015}
{Elbert} O.~D.,  {Bullock} J.~S.,  {Garrison-Kimmel} S.,  {Rocha} M.,
  {O{\~n}orbe} J.,   {Peter} A.~H.~G.,  2015, \mn@doi [\mnras]
  {10.1093/mnras/stv1470}, \href
  {http://adsabs.harvard.edu/abs/2015MNRAS.453...29E} {453, 29}

\bibitem[\protect\citeauthoryear{{Ettori}, {Fabian}, {Allen}  \&
  {Johnstone}}{{Ettori} et~al.}{2002}]{ettori2002}
{Ettori} S.,  {Fabian} A.~C.,  {Allen} S.~W.,   {Johnstone} R.~M.,  2002,
  \mn@doi [\mnras] {10.1046/j.1365-8711.2002.05212.x}, \href
  {http://adsabs.harvard.edu/abs/2002MNRAS.331..635E} {331, 635}

\bibitem[\protect\citeauthoryear{{Fedeli}}{{Fedeli}}{2012}]{fedeli2012}
{Fedeli} C.,  2012, \mn@doi [\mnras] {10.1111/j.1365-2966.2012.21302.x}, \href
  {http://adsabs.harvard.edu/abs/2012MNRAS.424.1244F} {424, 1244}

\bibitem[\protect\citeauthoryear{{Finkbeiner} \& {Weiner}}{{Finkbeiner} \&
  {Weiner}}{2007}]{eDM}
{Finkbeiner} D.~P.,  {Weiner} N.,  2007, \mn@doi [Physical Review D]
  {10.1103/PhysRevD.76.083519}, \href
  {http://adsabs.harvard.edu/abs/2007PhRvD..76h3519F} {76, 083519}

\bibitem[\protect\citeauthoryear{{Flores} \& {Primack}}{{Flores} \&
  {Primack}}{1994}]{flores1994}
{Flores} R.~A.,  {Primack} J.~R.,  1994, \mn@doi [\apjl] {10.1086/187350},
  \href {http://adsabs.harvard.edu/abs/1994ApJ...427L...1F} {427, L1}

\bibitem[\protect\citeauthoryear{{Fry} et~al.,}{{Fry} et~al.}{2015}]{fry2015}
{Fry} A.~B.,  et~al., 2015, \mn@doi [\mnras] {10.1093/mnras/stv1330}, \href
  {http://adsabs.harvard.edu/abs/2015MNRAS.452.1468F} {452, 1468}

\bibitem[\protect\citeauthoryear{{Garrison-Kimmel}, {Rocha}, {Boylan-Kolchin},
  {Bullock}  \& {Lally}}{{Garrison-Kimmel} et~al.}{2013}]{garrison-kimmel2013}
{Garrison-Kimmel} S.,  {Rocha} M.,  {Boylan-Kolchin} M.,  {Bullock} J.~S.,
  {Lally} J.,  2013, \mn@doi [\mnras] {10.1093/mnras/stt984}, \href
  {http://adsabs.harvard.edu/abs/2013MNRAS.433.3539G} {433, 3539}

\bibitem[\protect\citeauthoryear{{Garrison-Kimmel}, {Boylan-Kolchin}, {Bullock}
   \& {Kirby}}{{Garrison-Kimmel} et~al.}{2014}]{garrison-kimmel2014}
{Garrison-Kimmel} S.,  {Boylan-Kolchin} M.,  {Bullock} J.~S.,   {Kirby} E.~N.,
  2014, \mn@doi [\mnras] {10.1093/mnras/stu1477}, \href
  {http://adsabs.harvard.edu/abs/2014MNRAS.444..222G} {444, 222}

\bibitem[\protect\citeauthoryear{{Gavazzi}}{{Gavazzi}}{2005}]{gavazzi2005}
{Gavazzi} R.,  2005, \mn@doi [\aap] {10.1051/0004-6361:20053166}, \href
  {http://adsabs.harvard.edu/abs/2005A%26A...443..793G} {443, 793}

\bibitem[\protect\citeauthoryear{{Governato} et~al.,}{{Governato}
  et~al.}{2012}]{governato2012}
{Governato} F.,  et~al., 2012, \mn@doi [MNRAS]
  {10.1111/j.1365-2966.2012.20696.x}, \href
  {http://adsabs.harvard.edu/abs/2012MNRAS.422.1231G} {422, 1231}

\bibitem[\protect\citeauthoryear{{Graham}, {Harnik}, {Rajendran}  \&
  {Saraswat}}{{Graham} et~al.}{2010}]{graham2010}
{Graham} P.~W.,  {Harnik} R.,  {Rajendran} S.,   {Saraswat} P.,  2010, \mn@doi
  [\prd] {10.1103/PhysRevD.82.063512}, \href
  {http://adsabs.harvard.edu/abs/2010PhRvD..82f3512G} {82, 063512}

\bibitem[\protect\citeauthoryear{{Harvey}, {Massey}, {Kitching}, {Taylor}  \&
  {Tittley}}{{Harvey} et~al.}{2015}]{harvey2015}
{Harvey} D.,  {Massey} R.,  {Kitching} T.,  {Taylor} A.,   {Tittley} E.,  2015,
  \mn@doi [Science] {10.1126/science.1261381}, \href
  {http://adsabs.harvard.edu/abs/2015Sci...347.1462H} {347, 1462}

\bibitem[\protect\citeauthoryear{{Hennawi}, {Dalal}, {Bode}  \&
  {Ostriker}}{{Hennawi} et~al.}{2007}]{hennawi2007}
{Hennawi} J.~F.,  {Dalal} N.,  {Bode} P.,   {Ostriker} J.~P.,  2007, \mn@doi
  [\apj] {10.1086/497362}, \href
  {http://adsabs.harvard.edu/abs/2007ApJ...654..714H} {654, 714}

\bibitem[\protect\citeauthoryear{{Hinshaw} et~al.,}{{Hinshaw}
  et~al.}{2013}]{hinshaw2013}
{Hinshaw} G.,  et~al., 2013, \mn@doi [\apjs] {10.1088/0067-0049/208/2/19},
  \href {http://adsabs.harvard.edu/abs/2013ApJS..208...19H} {208, 19}

\bibitem[\protect\citeauthoryear{{Kahlhoefer}, {Schmidt-Hoberg}, {Frandsen}  \&
  {Sarkar}}{{Kahlhoefer} et~al.}{2014}]{kahlhoefer2014}
{Kahlhoefer} F.,  {Schmidt-Hoberg} K.,  {Frandsen} M.~T.,   {Sarkar} S.,  2014,
  \mn@doi [\mnras] {10.1093/mnras/stt2097}, \href
  {http://adsabs.harvard.edu/abs/2014MNRAS.437.2865K} {437, 2865}

\bibitem[\protect\citeauthoryear{{Kamada}, {Kaplinghat}, {Pace}  \&
  {Yu}}{{Kamada} et~al.}{2017}]{kamada2017}
{Kamada} A.,  {Kaplinghat} M.,  {Pace} A.~B.,   {Yu} H.-B.,  2017, \mn@doi
  [Physical Review Letters] {10.1103/PhysRevLett.119.111102}, \href
  {http://adsabs.harvard.edu/abs/2017PhRvL.119k1102K} {119, 111102}

\bibitem[\protect\citeauthoryear{{Kaplinghat}, {Tulin}  \& {Yu}}{{Kaplinghat}
  et~al.}{2016}]{kaplinghat2016}
{Kaplinghat} M.,  {Tulin} S.,   {Yu} H.-B.,  2016, \mn@doi [Physical Review
  Letters] {10.1103/PhysRevLett.116.041302}, \href
  {http://adsabs.harvard.edu/abs/2016PhRvL.116d1302K} {116, 041302}

\bibitem[\protect\citeauthoryear{{Klypin}, {Kravtsov}, {Valenzuela}  \&
  {Prada}}{{Klypin} et~al.}{1999}]{klypin1999}
{Klypin} A.,  {Kravtsov} A.~V.,  {Valenzuela} O.,   {Prada} F.,  1999, \mn@doi
  [\apj] {10.1086/307643}, \href
  {http://adsabs.harvard.edu/abs/1999ApJ...522...82K} {522, 82}

\bibitem[\protect\citeauthoryear{{Klypin}, {Trujillo-Gomez}  \&
  {Primack}}{{Klypin} et~al.}{2011}]{klypin2011}
{Klypin} A.~A.,  {Trujillo-Gomez} S.,   {Primack} J.,  2011, \mn@doi [\apj]
  {10.1088/0004-637X/740/2/102}, \href
  {http://adsabs.harvard.edu/abs/2011ApJ...740..102K} {740, 102}

\bibitem[\protect\citeauthoryear{{Knollmann} \& {Knebe}}{{Knollmann} \&
  {Knebe}}{2009}]{knollmann2009}
{Knollmann} S.~R.,  {Knebe} A.,  2009, \mn@doi [\apjs]
  {10.1088/0067-0049/182/2/608}, \href
  {http://adsabs.harvard.edu/abs/2009ApJS..182..608K} {182, 608}

\bibitem[\protect\citeauthoryear{{Kuzio de Naray} \& {Kaufmann}}{{Kuzio de
  Naray} \& {Kaufmann}}{2011}]{kuziodenaray2011}
{Kuzio de Naray} R.,  {Kaufmann} T.,  2011, \mn@doi [\mnras]
  {10.1111/j.1365-2966.2011.18656.x}, \href
  {http://adsabs.harvard.edu/abs/2011MNRAS.414.3617K} {414, 3617}

\bibitem[\protect\citeauthoryear{{Kuzio de Naray} \& {Spekkens}}{{Kuzio de
  Naray} \& {Spekkens}}{2011}]{kuziodenaray2011b}
{Kuzio de Naray} R.,  {Spekkens} K.,  2011, \mn@doi [\apjl]
  {10.1088/2041-8205/741/2/L29}, \href
  {http://adsabs.harvard.edu/abs/2011ApJ...741L..29K} {741, L29}

\bibitem[\protect\citeauthoryear{{Loeb} \& {Weiner}}{{Loeb} \&
  {Weiner}}{2011}]{loeb2011}
{Loeb} A.,  {Weiner} N.,  2011, \mn@doi [Physical Review Letters]
  {10.1103/PhysRevLett.106.171302}, \href
  {http://adsabs.harvard.edu/abs/2011PhRvL.106q1302L} {106, 171302}

\bibitem[\protect\citeauthoryear{{Macci{\`o}}, {Dutton}  \& {van den
  Bosch}}{{Macci{\`o}} et~al.}{2008}]{maccio2008}
{Macci{\`o}} A.~V.,  {Dutton} A.~A.,   {van den Bosch} F.~C.,  2008, \mn@doi
  [\mnras] {10.1111/j.1365-2966.2008.14029.x}, \href
  {http://adsabs.harvard.edu/abs/2008MNRAS.391.1940M} {391, 1940}

\bibitem[\protect\citeauthoryear{{Madau}, {Shen}  \& {Governato}}{{Madau}
  et~al.}{2014}]{madau2014}
{Madau} P.,  {Shen} S.,   {Governato} F.,  2014, \mn@doi [\apjl]
  {10.1088/2041-8205/789/1/L17}, \href
  {http://adsabs.harvard.edu/abs/2014ApJ...789L..17M} {789, L17}

\bibitem[\protect\citeauthoryear{{Markevitch}, {Gonzalez}, {Clowe},
  {Vikhlinin}, {Forman}, {Jones}, {Murray}  \& {Tucker}}{{Markevitch}
  et~al.}{2004}]{markevitch2004}
{Markevitch} M.,  {Gonzalez} A.~H.,  {Clowe} D.,  {Vikhlinin} A.,  {Forman} W.,
   {Jones} C.,  {Murray} S.,   {Tucker} W.,  2004, \mn@doi [\apj]
  {10.1086/383178}, \href {http://adsabs.harvard.edu/abs/2004ApJ...606..819M}
  {606, 819}

\bibitem[\protect\citeauthoryear{{Martizzi}, {Teyssier}, {Moore}  \&
  {Wentz}}{{Martizzi} et~al.}{2012}]{martizzi2012}
{Martizzi} D.,  {Teyssier} R.,  {Moore} B.,   {Wentz} T.,  2012, \mn@doi
  [\mnras] {10.1111/j.1365-2966.2012.20879.x}, \href
  {http://adsabs.harvard.edu/abs/2012MNRAS.422.3081M} {422, 3081}

\bibitem[\protect\citeauthoryear{{Martizzi}, {Teyssier}  \& {Moore}}{{Martizzi}
  et~al.}{2013}]{martizzi2013}
{Martizzi} D.,  {Teyssier} R.,   {Moore} B.,  2013, \mn@doi [\mnras]
  {10.1093/mnras/stt297}, \href
  {http://adsabs.harvard.edu/abs/2013MNRAS.432.1947M} {432, 1947}

\bibitem[\protect\citeauthoryear{{McCullough} \& {Randall}}{{McCullough} \&
  {Randall}}{2013}]{mccullough2013}
{McCullough} M.,  {Randall} L.,  2013, \mn@doi [\jcap]
  {10.1088/1475-7516/2013/10/058}, \href
  {http://adsabs.harvard.edu/abs/2013JCAP...10..058M} {10, 058}

\bibitem[\protect\citeauthoryear{{McNamara} \& {Nulsen}}{{McNamara} \&
  {Nulsen}}{2007}]{mcnamara2007}
{McNamara} B.~R.,  {Nulsen} P.~E.~J.,  2007, \mn@doi [\araa]
  {10.1146/annurev.astro.45.051806.110625}, \href
  {http://adsabs.harvard.edu/abs/2007ARA%26A..45..117M} {45, 117}

\bibitem[\protect\citeauthoryear{{Medvedev}}{{Medvedev}}{2000}]{medvedev2000}
{Medvedev} M.~V.,  2000, ArXiv Astrophysics e-prints, \href
  {http://adsabs.harvard.edu/abs/2000astro.ph.10616M} {}

\bibitem[\protect\citeauthoryear{{Medvedev}}{{Medvedev}}{2001a}]{medvedev2001a}
{Medvedev} M.,  2001a, ArXiv Astrophysics e-prints, \href
  {http://adsabs.harvard.edu/abs/2001astro.ph..5156M} {}

\bibitem[\protect\citeauthoryear{{Medvedev}}{{Medvedev}}{2001b}]{medvedev2001b}
{Medvedev} M.~V.,  2001b, in American Astronomical Society Meeting Abstracts
  \#198. p.~815

\bibitem[\protect\citeauthoryear{{Medvedev}}{{Medvedev}}{2001c}]{medvedev2001c}
{Medvedev} M.~V.,  2001c, in {Wheeler} J.~C.,  {Martel} H.,  eds,  American
  Institute of Physics Conference Series Vol. 586, 20th Texas Symposium on
  relativistic astrophysics. pp 149--151 (\mn@eprint {} {astro-ph/0102400}),
  \mn@doi{10.1063/1.1419546}

\bibitem[\protect\citeauthoryear{{Medvedev}}{{Medvedev}}{2010a}]{medvedev2010}
{Medvedev} M.~V.,  2010a, preprint, \href
  {http://adsabs.harvard.edu/abs/2010arXiv1004.3377M} {} (\mn@eprint {arXiv}
  {1004.3377})

\bibitem[\protect\citeauthoryear{{Medvedev}}{{Medvedev}}{2010b}]{medvedev2010b}
{Medvedev} M.~V.,  2010b, \mn@doi [Journal of Physics A Mathematical General]
  {10.1088/1751-8113/43/37/372002}, \href
  {http://adsabs.harvard.edu/abs/2010JPhA...43K2002M} {43, 372002}

\bibitem[\protect\citeauthoryear{{Medvedev}}{{Medvedev}}{2014a}]{medvedev2014theo}
{Medvedev} M.~V.,  2014a, \mn@doi [\jcap] {10.1088/1475-7516/2014/06/063},
  \href {http://adsabs.harvard.edu/abs/2014JCAP...06..063M} {6, 063}

\bibitem[\protect\citeauthoryear{{Medvedev}}{{Medvedev}}{2014b}]{medvedev2014}
{Medvedev} M.~V.,  2014b, \mn@doi [Physical Review Letters]
  {10.1103/PhysRevLett.113.071303}, \href
  {http://adsabs.harvard.edu/abs/2014PhRvL.113g1303M} {113, 071303}

\bibitem[\protect\citeauthoryear{{Meneghetti} et~al.,}{{Meneghetti}
  et~al.}{2014}]{meneghetti2014}
{Meneghetti} M.,  et~al., 2014, \mn@doi [\apj] {10.1088/0004-637X/797/1/34},
  \href {http://adsabs.harvard.edu/abs/2014ApJ...797...34M} {797, 34}

\bibitem[\protect\citeauthoryear{{Merten} et~al.,}{{Merten}
  et~al.}{2011}]{merten2011}
{Merten} J.,  et~al., 2011, \mn@doi [\mnras]
  {10.1111/j.1365-2966.2011.19266.x}, \href
  {http://adsabs.harvard.edu/abs/2011MNRAS.417..333M} {417, 333}

\bibitem[\protect\citeauthoryear{{Merten} et~al.,}{{Merten}
  et~al.}{2015}]{merten2015}
{Merten} J.,  et~al., 2015, \mn@doi [\apj] {10.1088/0004-637X/806/1/4}, \href
  {http://adsabs.harvard.edu/abs/2015ApJ...806....4M} {806, 4}

\bibitem[\protect\citeauthoryear{{Moore}, {Ghigna}, {Governato}, {Lake},
  {Quinn}, {Stadel}  \& {Tozzi}}{{Moore} et~al.}{1999}]{moore1999}
{Moore} B.,  {Ghigna} S.,  {Governato} F.,  {Lake} G.,  {Quinn} T.,  {Stadel}
  J.,   {Tozzi} P.,  1999, \mn@doi [\apjl] {10.1086/312287}, \href
  {http://adsabs.harvard.edu/abs/1999ApJ...524L..19M} {524, L19}

\bibitem[\protect\citeauthoryear{{Navarro}, {Eke}  \& {Frenk}}{{Navarro}
  et~al.}{1996a}]{navarro1996b}
{Navarro} J.~F.,  {Eke} V.~R.,   {Frenk} C.~S.,  1996a, \mn@doi [\mnras]
  {10.1093/mnras/283.3.L72}, \href
  {http://adsabs.harvard.edu/abs/1996MNRAS.283L..72N} {283, L72}

\bibitem[\protect\citeauthoryear{{Navarro}, {Frenk}  \& {White}}{{Navarro}
  et~al.}{1996b}]{navarro1996}
{Navarro} J.~F.,  {Frenk} C.~S.,   {White} S.~D.~M.,  1996b, \mn@doi [\apj]
  {10.1086/177173}, \href {http://adsabs.harvard.edu/abs/1996ApJ...462..563N}
  {462, 563}

\bibitem[\protect\citeauthoryear{{Navarro}, {Frenk}  \& {White}}{{Navarro}
  et~al.}{1997}]{navarro1997}
{Navarro} J.~F.,  {Frenk} C.~S.,   {White} S.~D.~M.,  1997, \mn@doi [\apj]
  {10.1086/304888}, \href {http://adsabs.harvard.edu/abs/1997ApJ...490..493N}
  {490, 493}

\bibitem[\protect\citeauthoryear{{Newman}, {Treu}, {Ellis}  \& {Sand}}{{Newman}
  et~al.}{2011}]{newman2011}
{Newman} A.~B.,  {Treu} T.,  {Ellis} R.~S.,   {Sand} D.~J.,  2011, \mn@doi
  [\apj \ Letters] {10.1088/2041-8205/728/2/L39}, \href
  {http://adsabs.harvard.edu/abs/2011ApJ...728L..39N} {728, L39}

\bibitem[\protect\citeauthoryear{{Newman}, {Treu}, {Ellis}, {Sand}, {Nipoti},
  {Richard}  \& {Jullo}}{{Newman} et~al.}{2013a}]{newman2013a}
{Newman} A.~B.,  {Treu} T.,  {Ellis} R.~S.,  {Sand} D.~J.,  {Nipoti} C.,
  {Richard} J.,   {Jullo} E.,  2013a, \mn@doi [\apj]
  {10.1088/0004-637X/765/1/24}, \href
  {http://adsabs.harvard.edu/abs/2013ApJ...765...24N} {765, 24}

\bibitem[\protect\citeauthoryear{{Newman}, {Treu}, {Ellis}  \& {Sand}}{{Newman}
  et~al.}{2013b}]{newman2013b}
{Newman} A.~B.,  {Treu} T.,  {Ellis} R.~S.,   {Sand} D.~J.,  2013b, \mn@doi
  [\apj] {10.1088/0004-637X/765/1/25}, \href
  {http://adsabs.harvard.edu/abs/2013ApJ...765...25N} {765, 25}

\bibitem[\protect\citeauthoryear{{O{\~n}orbe}, {Boylan-Kolchin}, {Bullock},
  {Hopkins}, {Kere{\v s}}, {Faucher-Gigu{\`e}re}, {Quataert}  \&
  {Murray}}{{O{\~n}orbe} et~al.}{2015}]{Onorbe2015}
{O{\~n}orbe} J.,  {Boylan-Kolchin} M.,  {Bullock} J.~S.,  {Hopkins} P.~F.,
  {Kere{\v s}} D.,  {Faucher-Gigu{\`e}re} C.-A.,  {Quataert} E.,   {Murray} N.,
   2015, \mn@doi [\mnras] {10.1093/mnras/stv2072}, \href
  {http://adsabs.harvard.edu/abs/2015MNRAS.454.2092O} {454, 2092}

\bibitem[\protect\citeauthoryear{{Oguri} et~al.,}{{Oguri}
  et~al.}{2009}]{oguri2009}
{Oguri} M.,  et~al., 2009, \mn@doi [\apj] {10.1088/0004-637X/699/2/1038}, \href
  {http://adsabs.harvard.edu/abs/2009ApJ...699.1038O} {699, 1038}

\bibitem[\protect\citeauthoryear{{Oguri}, {Bayliss}, {Dahle}, {Sharon},
  {Gladders}, {Natarajan}, {Hennawi}  \& {Koester}}{{Oguri}
  et~al.}{2012}]{oguri2012}
{Oguri} M.,  {Bayliss} M.~B.,  {Dahle} H.,  {Sharon} K.,  {Gladders} M.~D.,
  {Natarajan} P.,  {Hennawi} J.~F.,   {Koester} B.~P.,  2012, \mn@doi [\mnras]
  {10.1111/j.1365-2966.2011.20248.x}, \href
  {http://adsabs.harvard.edu/abs/2012MNRAS.420.3213O} {420, 3213}

\bibitem[\protect\citeauthoryear{{Oh}, {de Blok}, {Brinks}, {Walter}  \&
  {Kennicutt}}{{Oh} et~al.}{2011}]{oh2011}
{Oh} S.-H.,  {de Blok} W.~J.~G.,  {Brinks} E.,  {Walter} F.,   {Kennicutt} Jr.
  R.~C.,  2011, \mn@doi [\aj] {10.1088/0004-6256/141/6/193}, \href
  {http://adsabs.harvard.edu/abs/2011AJ....141..193O} {141, 193}

\bibitem[\protect\citeauthoryear{{Oh} et~al.,}{{Oh} et~al.}{2015}]{oh2015}
{Oh} S.-H.,  et~al., 2015, \mn@doi [\aj] {10.1088/0004-6256/149/6/180}, \href
  {http://adsabs.harvard.edu/abs/2015AJ....149..180O} {149, 180}

\bibitem[\protect\citeauthoryear{{Okabe}, {Takada}, {Umetsu}, {Futamase}  \&
  {Smith}}{{Okabe} et~al.}{2010}]{okabe2010}
{Okabe} N.,  {Takada} M.,  {Umetsu} K.,  {Futamase} T.,   {Smith} G.~P.,  2010,
  \mn@doi [\pasj] {10.1093/pasj/62.3.811}, \href
  {http://adsabs.harvard.edu/abs/2010PASJ...62..811O} {62, 811}

\bibitem[\protect\citeauthoryear{{Papastergis} \& {Shankar}}{{Papastergis} \&
  {Shankar}}{2016}]{papastergis2016}
{Papastergis} E.,  {Shankar} F.,  2016, \mn@doi [\aap]
  {10.1051/0004-6361/201527854}, \href
  {http://adsabs.harvard.edu/abs/2016A%26A...591A..58P} {591, A58}

\bibitem[\protect\citeauthoryear{{Papastergis}, {Giovanelli}, {Haynes}  \&
  {Shankar}}{{Papastergis} et~al.}{2015}]{papastergis2015}
{Papastergis} E.,  {Giovanelli} R.,  {Haynes} M.~P.,   {Shankar} F.,  2015,
  \mn@doi [\aap] {10.1051/0004-6361/201424909}, \href
  {http://adsabs.harvard.edu/abs/2015A%26A...574A.113P} {574, A113}

\bibitem[\protect\citeauthoryear{{Parry}, {Eke}, {Frenk}  \& {Okamoto}}{{Parry}
  et~al.}{2012}]{parry2012}
{Parry} O.~H.,  {Eke} V.~R.,  {Frenk} C.~S.,   {Okamoto} T.,  2012, \mn@doi
  [\mnras] {10.1111/j.1365-2966.2011.19971.x}, \href
  {http://adsabs.harvard.edu/abs/2012MNRAS.419.3304P} {419, 3304}

\bibitem[\protect\citeauthoryear{{Peter}, {Rocha}, {Bullock}  \&
  {Kaplinghat}}{{Peter} et~al.}{2013}]{peter2013}
{Peter} A.~H.~G.,  {Rocha} M.,  {Bullock} J.~S.,   {Kaplinghat} M.,  2013,
  \mn@doi [\mnras] {10.1093/mnras/sts535}, \href
  {http://adsabs.harvard.edu/abs/2013MNRAS.430..105P} {430, 105}

\bibitem[\protect\citeauthoryear{{Planck Collaboration} et~al.,}{{Planck
  Collaboration} et~al.}{2015}]{planck2015}
{Planck Collaboration} et~al., 2015, preprint, \href
  {http://adsabs.harvard.edu/abs/2015arXiv150201589P} {} (\mn@eprint {arXiv}
  {1502.01589})

\bibitem[\protect\citeauthoryear{{Pontzen} \& {Governato}}{{Pontzen} \&
  {Governato}}{2012}]{pontzen2012}
{Pontzen} A.,  {Governato} F.,  2012, \mn@doi [\mnras]
  {10.1111/j.1365-2966.2012.20571.x}, \href
  {http://adsabs.harvard.edu/abs/2012MNRAS.421.3464P} {421, 3464}

\bibitem[\protect\citeauthoryear{{Postman} et~al.,}{{Postman}
  et~al.}{2012}]{postman2012}
{Postman} M.,  et~al., 2012, \mn@doi [\apjs] {10.1088/0067-0049/199/2/25},
  \href {http://adsabs.harvard.edu/abs/2012ApJS..199...25P} {199, 25}

\bibitem[\protect\citeauthoryear{{Prada}, {Klypin}, {Cuesta}, {Betancort-Rijo}
  \& {Primack}}{{Prada} et~al.}{2012}]{prada2012}
{Prada} F.,  {Klypin} A.~A.,  {Cuesta} A.~J.,  {Betancort-Rijo} J.~E.,
  {Primack} J.,  2012, \mn@doi [\mnras] {10.1111/j.1365-2966.2012.21007.x},
  \href {http://adsabs.harvard.edu/abs/2012MNRAS.423.3018P} {423, 3018}

\bibitem[\protect\citeauthoryear{{Randall}, {Markevitch}, {Clowe}, {Gonzalez}
  \& {Brada{\v c}}}{{Randall} et~al.}{2008}]{randall2008}
{Randall} S.~W.,  {Markevitch} M.,  {Clowe} D.,  {Gonzalez} A.~H.,   {Brada{\v
  c}} M.,  2008, \mn@doi [\apj] {10.1086/587859}, \href
  {http://adsabs.harvard.edu/abs/2008ApJ...679.1173R} {679, 1173}

\bibitem[\protect\citeauthoryear{{Read} \& {Gilmore}}{{Read} \&
  {Gilmore}}{2005}]{read2005}
{Read} J.~I.,  {Gilmore} G.,  2005, \mn@doi [\mnras]
  {10.1111/j.1365-2966.2004.08424.x}, \href
  {http://adsabs.harvard.edu/abs/2005MNRAS.356..107R} {356, 107}

\bibitem[\protect\citeauthoryear{{Read}, {Agertz}  \& {Collins}}{{Read}
  et~al.}{2016}]{read2016}
{Read} J.~I.,  {Agertz} O.,   {Collins} M.~L.~M.,  2016, \mn@doi [\mnras]
  {10.1093/mnras/stw713}, \href
  {http://adsabs.harvard.edu/abs/2016MNRAS.tmp..504R} {}

\bibitem[\protect\citeauthoryear{{Robertson}, {Massey}  \& {Eke}}{{Robertson}
  et~al.}{2017}]{robertson2017}
{Robertson} A.,  {Massey} R.,   {Eke} V.,  2017, \mn@doi [\mnras]
  {10.1093/mnras/stw2670}, \href
  {http://adsabs.harvard.edu/abs/2017MNRAS.465..569R} {465, 569}

\bibitem[\protect\citeauthoryear{{Rocha}, {Peter}, {Bullock}, {Kaplinghat},
  {Garrison-Kimmel}, {O{\~n}orbe}  \& {Moustakas}}{{Rocha}
  et~al.}{2013}]{rocha2013}
{Rocha} M.,  {Peter} A.~H.~G.,  {Bullock} J.~S.,  {Kaplinghat} M.,
  {Garrison-Kimmel} S.,  {O{\~n}orbe} J.,   {Moustakas} L.~A.,  2013, \mn@doi
  [MNRAS] {10.1093/mnras/sts514}, \href
  {http://adsabs.harvard.edu/abs/2013MNRAS.430...81R} {430, 81}

\bibitem[\protect\citeauthoryear{{S{\'a}nchez-Conde} \&
  {Prada}}{{S{\'a}nchez-Conde} \& {Prada}}{2014}]{sanchez-conde2014}
{S{\'a}nchez-Conde} M.~A.,  {Prada} F.,  2014, \mn@doi [\mnras]
  {10.1093/mnras/stu1014}, \href
  {http://adsabs.harvard.edu/abs/2014MNRAS.442.2271S} {442, 2271}

\bibitem[\protect\citeauthoryear{{Sand}, {Treu}, {Smith}  \& {Ellis}}{{Sand}
  et~al.}{2004}]{sand2004}
{Sand} D.~J.,  {Treu} T.,  {Smith} G.~P.,   {Ellis} R.~S.,  2004, \mn@doi
  [\apj] {10.1086/382146}, \href
  {http://adsabs.harvard.edu/abs/2004ApJ...604...88S} {604, 88}

\bibitem[\protect\citeauthoryear{{Sawala}, {Frenk}, {Crain}, {Jenkins},
  {Schaye}, {Theuns}  \& {Zavala}}{{Sawala} et~al.}{2013}]{sawala2013}
{Sawala} T.,  {Frenk} C.~S.,  {Crain} R.~A.,  {Jenkins} A.,  {Schaye} J.,
  {Theuns} T.,   {Zavala} J.,  2013, \mn@doi [\mnras] {10.1093/mnras/stt259},
  \href {http://adsabs.harvard.edu/abs/2013MNRAS.431.1366S} {431, 1366}

\bibitem[\protect\citeauthoryear{{Sawala} et~al.,}{{Sawala}
  et~al.}{2016}]{sawala2016}
{Sawala} T.,  et~al., 2016, \mn@doi [\mnras] {10.1093/mnras/stw145}, \href
  {http://adsabs.harvard.edu/abs/2016MNRAS.457.1931S} {457, 1931}

\bibitem[\protect\citeauthoryear{{Sawala}, {Pihajoki}, {Johansson}, {Frenk},
  {Navarro}, {Oman}  \& {White}}{{Sawala} et~al.}{2017}]{sawala2017}
{Sawala} T.,  {Pihajoki} P.,  {Johansson} P.~H.,  {Frenk} C.~S.,  {Navarro}
  J.~F.,  {Oman} K.~A.,   {White} S.~D.~M.,  2017, \mn@doi [\mnras]
  {10.1093/mnras/stx360}, \href
  {http://adsabs.harvard.edu/abs/2017MNRAS.467.4383S} {467, 4383}

\bibitem[\protect\citeauthoryear{{Schaller} et~al.,}{{Schaller}
  et~al.}{2015}]{schaller2015}
{Schaller} M.,  et~al., 2015, \mn@doi [\mnras] {10.1093/mnras/stv1341}, \href
  {http://adsabs.harvard.edu/abs/2015MNRAS.452..343S} {452, 343}

\bibitem[\protect\citeauthoryear{{Schmidt} \& {Allen}}{{Schmidt} \&
  {Allen}}{2007}]{schmidt2007}
{Schmidt} R.~W.,  {Allen} S.~W.,  2007, \mn@doi [\mnras]
  {10.1111/j.1365-2966.2007.11928.x}, \href
  {http://adsabs.harvard.edu/abs/2007MNRAS.379..209S} {379, 209}

\bibitem[\protect\citeauthoryear{{Sereno}, {Giocoli}, {Ettori}  \&
  {Moscardini}}{{Sereno} et~al.}{2015}]{sereno2015}
{Sereno} M.,  {Giocoli} C.,  {Ettori} S.,   {Moscardini} L.,  2015, \mn@doi
  [\mnras] {10.1093/mnras/stv416}, \href
  {http://adsabs.harvard.edu/abs/2015MNRAS.449.2024S} {449, 2024}

\bibitem[\protect\citeauthoryear{{Simpson}, {Bryan}, {Johnston}, {Smith}, {Mac
  Low}, {Sharma}  \& {Tumlinson}}{{Simpson} et~al.}{2013}]{simpson2013}
{Simpson} C.~M.,  {Bryan} G.~L.,  {Johnston} K.~V.,  {Smith} B.~D.,  {Mac Low}
  M.-M.,  {Sharma} S.,   {Tumlinson} J.,  2013, \mn@doi [\mnras]
  {10.1093/mnras/stt474}, \href
  {http://adsabs.harvard.edu/abs/2013MNRAS.432.1989S} {432, 1989}

\bibitem[\protect\citeauthoryear{{Smith} \& {Weiner}}{{Smith} \&
  {Weiner}}{2001}]{iDM}
{Smith} D.,  {Weiner} N.,  2001, \mn@doi [Physical Review D]
  {10.1103/PhysRevD.64.043502}, \href
  {http://adsabs.harvard.edu/abs/2001PhRvD..64d3502S} {64, 043502}

\bibitem[\protect\citeauthoryear{{Spergel} \& {Steinhardt}}{{Spergel} \&
  {Steinhardt}}{2000}]{spergel2000}
{Spergel} D.~N.,  {Steinhardt} P.~J.,  2000, \mn@doi [Physical Review Letters]
  {10.1103/PhysRevLett.84.3760}, \href
  {http://adsabs.harvard.edu/abs/2000PhRvL..84.3760S} {84, 3760}

\bibitem[\protect\citeauthoryear{{Springel}}{{Springel}}{2005}]{springel2005}
{Springel} V.,  2005, \mn@doi [\mnras] {10.1111/j.1365-2966.2005.09655.x},
  \href {http://adsabs.harvard.edu/abs/2005MNRAS.364.1105S} {364, 1105}

\bibitem[\protect\citeauthoryear{{Springel} et~al.,}{{Springel}
  et~al.}{2005}]{springel2005_millennium}
{Springel} V.,  et~al., 2005, \mn@doi [Nature] {10.1038/nature03597}, \href
  {https://ui.adsabs.harvard.edu/abs/2005Natur.435..629S} {435, 629}

\bibitem[\protect\citeauthoryear{{Springel} et~al.,}{{Springel}
  et~al.}{2008}]{springel2008}
{Springel} V.,  et~al., 2008, \mn@doi [\mnras]
  {10.1111/j.1365-2966.2008.14066.x}, \href
  {http://adsabs.harvard.edu/abs/2008MNRAS.391.1685S} {391, 1685}

\bibitem[\protect\citeauthoryear{{Stadel}, {Potter}, {Moore}, {Diemand},
  {Madau}, {Zemp}, {Kuhlen}  \& {Quilis}}{{Stadel} et~al.}{2009}]{stadel2009}
{Stadel} J.,  {Potter} D.,  {Moore} B.,  {Diemand} J.,  {Madau} P.,  {Zemp} M.,
   {Kuhlen} M.,   {Quilis} V.,  2009, \mn@doi [\mnras]
  {10.1111/j.1745-3933.2009.00699.x}, \href
  {http://adsabs.harvard.edu/abs/2009MNRAS.398L..21S} {398, L21}

\bibitem[\protect\citeauthoryear{{Strigari}, {Bullock}, {Kaplinghat}, {Simon},
  {Geha}, {Willman}  \& {Walker}}{{Strigari} et~al.}{2008}]{strigari2008b}
{Strigari} L.~E.,  {Bullock} J.~S.,  {Kaplinghat} M.,  {Simon} J.~D.,  {Geha}
  M.,  {Willman} B.,   {Walker} M.~G.,  2008, \mn@doi [\nat]
  {10.1038/nature07222}, \href
  {http://adsabs.harvard.edu/abs/2008Natur.454.1096S} {454, 1096}

\bibitem[\protect\citeauthoryear{{Swaters}, {Madore}, {van den Bosch}  \&
  {Balcells}}{{Swaters} et~al.}{2003}]{swaters2003}
{Swaters} R.~A.,  {Madore} B.~F.,  {van den Bosch} F.~C.,   {Balcells} M.,
  2003, \mn@doi [\apj] {10.1086/345426}, \href
  {http://adsabs.harvard.edu/abs/2003ApJ...583..732S} {583, 732}

\bibitem[\protect\citeauthoryear{{Tegmark} et~al.,}{{Tegmark}
  et~al.}{2006}]{tegmark2006}
{Tegmark} M.,  et~al., 2006, \mn@doi [\prd] {10.1103/PhysRevD.74.123507}, \href
  {http://adsabs.harvard.edu/abs/2006PhRvD..74l3507T} {74, 123507}

\bibitem[\protect\citeauthoryear{{Teyssier}, {Pontzen}, {Dubois}  \&
  {Read}}{{Teyssier} et~al.}{2013}]{teyssier2013}
{Teyssier} R.,  {Pontzen} A.,  {Dubois} Y.,   {Read} J.~I.,  2013, \mn@doi
  [\mnras] {10.1093/mnras/sts563}, \href
  {http://adsabs.harvard.edu/abs/2013MNRAS.429.3068T} {429, 3068}

\bibitem[\protect\citeauthoryear{{Todoroki} \& {Medvedev}}{{Todoroki} \&
  {Medvedev}}{2019a}]{todoroki2019a}
{Todoroki} K.,  {Medvedev} M.~V.,  2019a, \mn@doi [\mnras]
  {10.1093/mnras/sty3401}, \href
  {https://ui.adsabs.harvard.edu/abs/2019MNRAS.483.3983T} {483, 3983}

\bibitem[\protect\citeauthoryear{{Todoroki} \& {Medvedev}}{{Todoroki} \&
  {Medvedev}}{2019b}]{todoroki2019b}
{Todoroki} K.,  {Medvedev} M.~V.,  2019b, \mn@doi [\mnras]
  {10.1093/mnras/sty3353}, \href
  {https://ui.adsabs.harvard.edu/abs/2019MNRAS.483.4004T} {483, 4004}

\bibitem[\protect\citeauthoryear{{Tollerud}, {Boylan-Kolchin}  \&
  {Bullock}}{{Tollerud} et~al.}{2014}]{tollerud2014}
{Tollerud} E.~J.,  {Boylan-Kolchin} M.,   {Bullock} J.~S.,  2014, \mn@doi
  [\mnras] {10.1093/mnras/stu474}, \href
  {http://adsabs.harvard.edu/abs/2014MNRAS.440.3511T} {440, 3511}

\bibitem[\protect\citeauthoryear{{Tollet} et~al.,}{{Tollet}
  et~al.}{2016}]{tollet2016}
{Tollet} E.,  et~al., 2016, \mn@doi [\mnras] {10.1093/mnras/stv2856}, \href
  {http://adsabs.harvard.edu/abs/2016MNRAS.456.3542T} {456, 3542}

\bibitem[\protect\citeauthoryear{{Torri}, {Meneghetti}, {Bartelmann},
  {Moscardini}, {Rasia}  \& {Tormen}}{{Torri} et~al.}{2004}]{torri2004}
{Torri} E.,  {Meneghetti} M.,  {Bartelmann} M.,  {Moscardini} L.,  {Rasia} E.,
   {Tormen} G.,  2004, \mn@doi [\mnras] {10.1111/j.1365-2966.2004.07508.x},
  \href {http://adsabs.harvard.edu/abs/2004MNRAS.349..476T} {349, 476}

\bibitem[\protect\citeauthoryear{{Trujillo-Gomez}, {Schneider}, {Papastergis},
  {Reed}  \& {Lake}}{{Trujillo-Gomez} et~al.}{2016}]{trujillo2016}
{Trujillo-Gomez} S.,  {Schneider} A.,  {Papastergis} E.,  {Reed} D.~S.,
  {Lake} G.,  2016, preprint, \href
  {http://adsabs.harvard.edu/abs/2016arXiv161009335T} {} (\mn@eprint {arXiv}
  {1610.09335})

\bibitem[\protect\citeauthoryear{{Umetsu}, {Zitrin}, {Gruen}, {Merten},
  {Donahue}  \& {Postman}}{{Umetsu} et~al.}{2016}]{umetsu2016}
{Umetsu} K.,  {Zitrin} A.,  {Gruen} D.,  {Merten} J.,  {Donahue} M.,
  {Postman} M.,  2016, \mn@doi [\apj] {10.3847/0004-637X/821/2/116}, \href
  {http://adsabs.harvard.edu/abs/2016ApJ...821..116U} {821, 116}

\bibitem[\protect\citeauthoryear{{Vogelsberger}, {Zavala}  \&
  {Loeb}}{{Vogelsberger} et~al.}{2012}]{vogelsberger2012}
{Vogelsberger} M.,  {Zavala} J.,   {Loeb} A.,  2012, \mn@doi [\mnras]
  {10.1111/j.1365-2966.2012.21182.x}, \href
  {http://adsabs.harvard.edu/abs/2012MNRAS.423.3740V} {423, 3740}

\bibitem[\protect\citeauthoryear{{Vogelsberger}, {Zavala}, {Simpson}  \&
  {Jenkins}}{{Vogelsberger} et~al.}{2014}]{vogelsberger2014b}
{Vogelsberger} M.,  {Zavala} J.,  {Simpson} C.,   {Jenkins} A.,  2014, \mn@doi
  [MNRAS] {10.1093/mnras/stu1713}, \href
  {http://adsabs.harvard.edu/abs/2014MNRAS.444.3684V} {444, 3684}

\bibitem[\protect\citeauthoryear{{Vogelsberger}, {Zavala}, {Cyr-Racine},
  {Pfrommer}, {Bringmann}  \& {Sigurdson}}{{Vogelsberger}
  et~al.}{2016}]{vogelsberger2016}
{Vogelsberger} M.,  {Zavala} J.,  {Cyr-Racine} F.-Y.,  {Pfrommer} C.,
  {Bringmann} T.,   {Sigurdson} K.,  2016, \mn@doi [\mnras]
  {10.1093/mnras/stw1076}, \href
  {http://adsabs.harvard.edu/abs/2016MNRAS.460.1399V} {460, 1399}

\bibitem[\protect\citeauthoryear{{Vogelsberger}, {Zavala}, {Schutz}  \&
  {Slatyer}}{{Vogelsberger} et~al.}{2019}]{vogelsberger2019}
{Vogelsberger} M.,  {Zavala} J.,  {Schutz} K.,   {Slatyer} T.~R.,  2019,
  \mn@doi [\mnras] {10.1093/mnras/stz340}, \href
  {https://ui.adsabs.harvard.edu/abs/2019MNRAS.484.5437V} {484, 5437}

\bibitem[\protect\citeauthoryear{{Walker} \& {Pe{\~n}arrubia}}{{Walker} \&
  {Pe{\~n}arrubia}}{2011}]{walker2011}
{Walker} M.~G.,  {Pe{\~n}arrubia} J.,  2011, \mn@doi [\apj]
  {10.1088/0004-637X/742/1/20}, \href
  {http://adsabs.harvard.edu/abs/2011ApJ...742...20W} {742, 20}

\bibitem[\protect\citeauthoryear{{Wetzel}, {Hopkins}, {Kim},
  {Faucher-Gigu{\`e}re}, {Kere{\v s}}  \& {Quataert}}{{Wetzel}
  et~al.}{2016}]{wetzel2016}
{Wetzel} A.~R.,  {Hopkins} P.~F.,  {Kim} J.-h.,  {Faucher-Gigu{\`e}re} C.-A.,
  {Kere{\v s}} D.,   {Quataert} E.,  2016, \mn@doi [\apjl]
  {10.3847/2041-8205/827/2/L23}, \href
  {http://adsabs.harvard.edu/abs/2016ApJ...827L..23W} {827, L23}

\bibitem[\protect\citeauthoryear{{White}, {van Waerbeke}  \& {Mackey}}{{White}
  et~al.}{2002}]{white2002}
{White} M.,  {van Waerbeke} L.,   {Mackey} J.,  2002, \mn@doi [\apj]
  {10.1086/341351}, \href {http://adsabs.harvard.edu/abs/2002ApJ...575..640W}
  {575, 640}

\bibitem[\protect\citeauthoryear{{Zavala}, {Vogelsberger}  \&
  {Walker}}{{Zavala} et~al.}{2013}]{zavala2013}
{Zavala} J.,  {Vogelsberger} M.,   {Walker} M.~G.,  2013, \mn@doi [\mnras]
  {10.1093/mnrasl/sls053}, \href
  {http://adsabs.harvard.edu/abs/2013MNRAS.431L..20Z} {431, L20}

\bibitem[\protect\citeauthoryear{{de Blok}}{{de Blok}}{2010}]{deblok2010}
{de Blok} W.~J.~G.,  2010, \mn@doi [Advances in Astronomy]
  {10.1155/2010/789293}, \href
  {http://adsabs.harvard.edu/abs/2010AdAst2010E...5D} {2010, 789293}

\bibitem[\protect\citeauthoryear{{de Blok}, {Walter}, {Brinks}, {Trachternach},
  {Oh}  \& {Kennicutt}}{{de Blok} et~al.}{2008}]{deblok2008}
{de Blok} W.~J.~G.,  {Walter} F.,  {Brinks} E.,  {Trachternach} C.,  {Oh}
  S.-H.,   {Kennicutt} Jr. R.~C.,  2008, \mn@doi [\aj]
  {10.1088/0004-6256/136/6/2648}, \href
  {http://adsabs.harvard.edu/abs/2008AJ....136.2648D} {136, 2648}

\bibitem[\protect\citeauthoryear{{di Cintio}, {Knebe}, {Libeskind}, {Yepes},
  {Gottl{\"o}ber}  \& {Hoffman}}{{di Cintio} et~al.}{2011}]{dicintio2011}
{di Cintio} A.,  {Knebe} A.,  {Libeskind} N.~I.,  {Yepes} G.,  {Gottl{\"o}ber}
  S.,   {Hoffman} Y.,  2011, \mn@doi [\mnras]
  {10.1111/j.1745-3933.2011.01123.x}, \href
  {http://adsabs.harvard.edu/abs/2011MNRAS.417L..74D} {417, L74}

\bibitem[\protect\citeauthoryear{{van der Wel} et~al.,}{{van der Wel}
  et~al.}{2011}]{vanderwel2011}
{van der Wel} A.,  et~al., 2011, \mn@doi [\apj] {10.1088/0004-637X/742/2/111},
  \href {http://adsabs.harvard.edu/abs/2011ApJ...742..111V} {742, 111}

\makeatother
\end{thebibliography}

\appendix

\label{lastpage}

\end{document}